\documentclass[numbers,10pt]{sigplanconf}

\usepackage{amsmath}

\usepackage{url}

\usepackage{balance}

\usepackage{amsmath}
\usepackage{amssymb}
\usepackage{multicol}
\usepackage{verbatim}
\usepackage{enumitem}
\usepackage{graphicx}
\usepackage{algorithm}
\usepackage{algpseudocode}
\usepackage{mathtools}
\usepackage{multirow}
\usepackage{bm}
\usepackage{nicefrac}

\usepackage{etoolbox}
\usepackage{float}
\newfloat{algorithm}{t}{lop}

\usepackage{moresize}

\usepackage[usenames,dvipsnames]{color}
\usepackage{pifont}
\newcommand{\cmark}{\text{\color{Green}\ding{51}}}
\newcommand{\xmark}{\text{\color{Red}\ding{55}}}
\newcommand{\texttta}[1]{\texttt{#1}}
\newcommand{\textttb}[1]{\texttt{\color{Cerulean}#1}}
\newcommand{\blue}[1]{{\color{Cerulean}#1}}

\usepackage{xspace}
\newcommand{\toolname}{{\sc Glade}\xspace}

\floatname{algorithm}{\small Algorithm}

\input{macros.tex}

\newcommand{\sss}{\hspace{0.1in}}

\makeatletter
\newcommand*{\da@rightarrow}{\mathchar"0\hexnumber@\symAMSa 4B }
\newcommand*{\da@leftarrow}{\mathchar"0\hexnumber@\symAMSa 4C }
\newcommand*{\xdashrightarrow}[2][]{%
  \mathrel{%
    \mathpalette{\da@xarrow{#1}{#2}{}\da@rightarrow{\,}{}}{}%
  }%
}
\newcommand{\xdashleftarrow}[2][]{%
  \mathrel{%
    \mathpalette{\da@xarrow{#1}{#2}\da@leftarrow{}{}{\,}}{}%
  }%
}
\newcommand*{\da@xarrow}[7]{%
  \sbox0{$\ifx#7\scriptstyle\scriptscriptstyle\else\scriptstyle\fi#5#1#6\m@th$}%
  \sbox2{$\ifx#7\scriptstyle\scriptscriptstyle\else\scriptstyle\fi#5#2#6\m@th$}%
  \sbox4{$#7\dabar@\m@th$}%
  \dimen@=\wd0 %
  \ifdim\wd2 >\dimen@
    \dimen@=\wd2 %   
  \fi
  \count@=2 %
  \def\da@bars{\dabar@\dabar@}%
  \@whiledim\count@\wd4<\dimen@\do{%
    \advance\count@\@ne
    \expandafter\def\expandafter\da@bars\expandafter{%
      \da@bars
      \dabar@ 
    }%
  }%  
  \mathrel{#3}%
  \mathrel{%   
    \mathop{\da@bars}\limits
    \ifx\\#1\\%
    \else
      _{\copy0}%
    \fi
    \ifx\\#2\\%
    \else
      ^{\copy2}%
    \fi
  }%   
  \mathrel{#4}%
}
\makeatother

\begin{document}

\toappear{}

\setlength{\pdfpageheight}{\paperheight}
\setlength{\pdfpagewidth}{\paperwidth}

%\conferenceinfo{CONF 'yy}{Month d--d, 20yy, City, ST, Country}
%\copyrightyear{20yy}
%\copyrightdata{978-1-nnnn-nnnn-n/yy/mm}
%\copyrightdoi{nnnnnnn.nnnnnnn}

% Uncomment the publication rights you want to use.
%\publicationrights{transferred}
%\publicationrights{licensed}     % this is the default
%\publicationrights{author-pays}

%\titlebanner{banner above paper title}        % These are ignored unless
%\preprintfooter{grammar synthesis for fuzz testing}   % 'preprint' option specified.

\title{Synthesizing Program Input Grammars}
%\subtitle{Subtitle Text, if any}

\authorinfo{Osbert Bastani}
           {Stanford University, USA}
           {obastani@cs.stanford.edu}
\authorinfo{Rahul Sharma}
           {Microsoft Research, India}
           {rahsha@microsoft.com}
\authorinfo{Alex Aiken}
           {Stanford University, USA}
           {aiken@cs.stanford.edu}
\authorinfo{Percy Liang}
           {Stanford University, USA}
           {pliang@cs.stanford.edu}

\maketitle

\begin{abstract}
We present an algorithm for synthesizing a context-free grammar encoding the language of valid program inputs from a set of input examples and blackbox access to the program. Our algorithm addresses shortcomings of existing grammar inference algorithms, which both severely overgeneralize and are prohibitively slow. Our implementation, \toolname, leverages the grammar synthesized by our algorithm to fuzz test programs with structured inputs. We show that \toolname substantially increases the incremental coverage on valid inputs compared to two baseline fuzzers. 
\end{abstract}

\begin{CCSXML}
<ccs2012>
<concept>
<concept_id>10003752.10010124.10010138.10010143</concept_id>
<concept_desc>Theory of computation~Program analysis</concept_desc>
<concept_significance>300</concept_significance>
</concept>
</ccs2012>
\end{CCSXML}

\ccsdesc[300]{Theory of computation~Program analysis}
%\category{F.3.2}{Semantics of Programming Languages}{Program analysis}

\keywords
grammar synthesis; fuzzing

\section{Introduction}
\label{sec:intro}

Documentation of program input formats, if available in a machine-readable form, can significantly aid many software analysis tools. However, such documentation is often poor; for example, the specifications of Flex~\cite{Flex} and Bison~\cite{Bison} input syntaxes are limited to informal documentation. Even when detailed specifications are available, they are often not in a machine-readable form; for example, the specification for ECMAScript 6 syntax is 20 pages in Annex A of~\cite{EcmaScript}, and the specification for Java class files is 268 pages in Chapter 4 of~\cite{Java}.

In this paper, we study the problem of automatically synthesizing grammars representing program input languages. Such a grammar synthesis algorithm has many potential applications. Our primary motivation is the possibility of using synthesized grammars with grammar-based fuzzers~\cite{majumdar2007directed,godefroid2008grammar,holler2012fuzzing}. For example, such inputs can be used to find bugs in real-world programs~\cite{purdom1972sentence,miller1990empirical,godefroid2008automated,yang2011finding}, learn abstractions~\cite{naik2012abstractions}, predict performance~\cite{huang2010predicting}, and aid dynamic analysis~\cite{nethercote2007valgrind}. Beyond fuzzing, a grammar synthesis algorithm could be used to reverse engineer input formats~\cite{hoschele2016mining}, in particular, network protocol message formats can help security analysts discover vulnerabilities in network programs~\cite{caballero2007polyglot,wondracek2008automatic,lin2008deriving,lin2010reverse}. Synthesized grammars could also be used to whitelist program inputs, thereby preventing exploits~\cite{rinard2003acceptability,su2006essence,rinard2007living}.

Approaches to synthesizing program input grammars typically examine executions of the program, and then generalize these observations to a representation of valid inputs. These approaches can be either \emph{whitebox} or \emph{blackbox}. Whitebox approaches assume that the program code is available for analysis and instrumentation, for example, using dynamic taint analysis~\cite{hoschele2016mining}. Such an approach is difficult when only the program binaries are available or when parts of the code (e.g., libraries) are missing. Furthermore, these techniques often require program-specific configuration or tuning, and may be affected by the structure of the code. We consider the blackbox setting, where we only require the ability to execute the program on a given input and observe its corresponding output. Since the algorithm does not examine the program's code, its performance depends only on the language of valid inputs, and not on implementation details.

A number of existing language inference algorithms can be adapted to this setting~\cite{de2010grammatical}. However, we found them to be unsuitable for synthesizing program input grammars. In particular, $L$-Star~\cite{angluin1987learning} and RPNI~\cite{oncina1992identifying}, the most widely studied algorithms~\cite{vardhan2004learning,cho2011mace,giannakopoulou2012symbolic,botinvcan2013sigma,choi2013guided}, were unable to learn or approximate even simple input languages such as XML, and furthermore do not scale even to small sets of seed inputs. Surprisingly, we found that $L$-Star and RPNI perform poorly even on the class of regular languages they target.

The problem with these algorithms is that despite having theoretical guarantees, they depend on assumptions that do not hold in the setting of learning program input grammars. For example, they typically avoid overgeneralizing by relying on an ``oracle'' to provide negative examples that are used by the algorithm to identify and remove overly general portions of the language. However, these oracles are not available in our setting---e.g., $L$-Star obtains such examples from an equivalence oracle, and RPNI obtains them ``in the limit''. They likewise assume that positive examples exercising all interesting behaviors are provided by this oracle. In our setting, the needed positive and negative examples are difficult to find, and existing algorithms consistently overgeneralize (e.g., return $\Sigma^*$) or undergeneralize (e.g., return $\emptyset$). Additionally, despite having polynomial running time, they can be very slow on our problem instances. To the best of our knowledge, other existing grammar inference algorithms are either impractical~\cite{lee1996learning,de2010grammatical} or make assumptions similar to $L$-Star and RPNI~\cite{ishizaka1990polynomial}.

This paper presents the first practical algorithm for synthesizing program input grammars in the blackbox setting. Our algorithm synthesizes a context-free grammar $\hat{C}$ encoding the language $L_*$ of valid program inputs, given
\begin{itemize}
\item A small set of \emph{seed inputs} $E_{\text{in}}\subseteq L_*$ (i.e., examples of valid inputs). Typically, seed inputs are readily available---in our evaluation, we use small test suites that come with programs or examples from documentation.
\item Blackbox access to the program executable to answer \emph{membership queries} (i.e., whether a given input is valid).
\end{itemize}

Our algorithm adopts a high-level design commonly used by language learning algorithms (e.g., RPNI)---it starts with the language containing exactly the given positive examples, and then incrementally generalizes this language, using negative examples to avoid overgeneralizing. Our algorithm avoids the shortcomings of existing algorithms in two ways:
\begin{itemize}
\item It considers a much richer set of potential generalizations, which addresses the issue of omitted positive examples.
\item It generates negative examples on the fly to avoid overgeneralizing, which addresses the issue of omitted negative examples.
\end{itemize}

In particular, our algorithm constructs a series of increasingly general languages using \emph{generalization steps}. Each step first proposes a number of candidate languages that generalize the current language, and then uses carefully crafted membership queries to reject candidates that overgeneralize. Our algorithm considers candidates that (i) add repetition and alternation constructs characteristic of regular expressions, (ii) induce recursive productions characteristic of context-free grammars, in particular, parentheses matching grammars, and (iii) generalize constants in the grammar.

We implement our approach in a tool called \toolname,\footnote{\toolname stands for Grammar Learning for AutomateD Execution, and is available at \url{https://github.com/obastani/glade}.}. We conduct an extensive empirical evaluation of \toolname (Section~\ref{sec:exp}), and show that \toolname substantially outperforms both $L$-Star and RPNI, even when restricted to synthesizing regular expressions. Furthermore, we show that \toolname successfully synthesizes input grammars for real programs, which can be used to fuzz test those programs. In particular, \toolname automatically synthesizes a program input grammar, and then uses the synthesized grammar in conjunction with a standard grammar-based fuzzer (described in Section~\ref{sec:expfuzz}) to generate new test inputs. Many fuzzing applications require valid inputs, for example, differential testing~\cite{yang2011finding}. We show that when restricted to generating valid inputs, \toolname increases line coverage compared to both a na\"{i}ve fuzzer and a production fuzzer afl-fuzz~\cite{zalewski2015afl}. Our contributions are:
\begin{itemize}
\item We introduce an algorithm for synthesizing program input grammars from seed inputs and blackbox program access (Section~\ref{sec:algo}). Our algorithm first learns regular properties such as repetitions and alternations (Section~\ref{sec:phaseone}), and then learns recursive productions characteristic of matching parentheses grammars (Section~\ref{sec:phasetwo}).
\item We implement our grammar synthesis algorithm in a tool called \toolname, and show that \toolname outperforms two widely studied language learning algorithms, $L$-Star and RPNI, in our application domain (Section~\ref{sec:explearning}).
\item We use \toolname to fuzz test programs, showing that it increases the number of newly covered lines of code using valid inputs by up to 6$\times$ compared to two baseline fuzzers (Section~\ref{sec:expfuzz}).
\end{itemize}

\section{Problem Formulation}
\label{sec:prob}

Suppose we are given a program that takes inputs in $\Sigma^*$, where $\Sigma$ is the input alphabet (e.g., ASCII characters). We let $L_*\subseteq\Sigma^*$ denote the \emph{target language} of valid program inputs; typically, $L_*$ is a highly structured subset of $\Sigma^*$. Our goal is to synthesize a language $\hat{L}$ approximating $L_*$ from blackbox program access and seed inputs $E_{\text{in}}\subseteq L_*$. We represent blackbox program access as an oracle $\mathcal{O}$ such that $\mathcal{O}(\alpha)=\mathbb{I}[\alpha\in L_*]$ (here, $\mathbb{I}$ is the indicator function, so $\mathbb{I}[\mathcal{C}]$ is $1$ if $\mathcal{C}$ is true and $0$ otherwise). In particular, we run the program on input $\alpha\in\Sigma^*$, and conclude that $\alpha$ is a valid input (i.e., $\alpha\in L_*$) if the program does not print an error message. Access to the oracle is crucial to avoid overgeneralizing, e.g., rejecting $\hat{L}=\Sigma^*$, whereas the seed inputs give a starting point from which to generalize.

As a running example, suppose the program input language is the XML-like grammar $C_{\text{XML}}$ shown in Figure~\ref{fig:xml}. We use $+$ to denote alternations and $*$ (the Kleene star) to denote repetitions. Terminals that are part of regular expressions or context-free grammars are highlighted in blue. Given seed input $\alpha_{\text{XML}}$ and oracle $\mathcal{O}_{\text{XML}}$, our goal is to synthesize a language $\hat{L}$ approximating $L_*=\mathcal{L}(C_{\text{XML}})$.

\begin{figure}
\small
\begin{itemize}
\item Target language $\mathcal{L}(C_{\text{XML}})$, where the context-free grammar $C_{\text{XML}}$ has terminals $\Sigma_{\text{XML}}=\{\textttb{a},...,\textttb{z},\textttb{<},\textttb{>},\textttb{/}\}$, start symbol $A_{\text{XML}}$, and production
\begin{align*}
A_{\text{XML}}\to(\textttb{a}+...+\textttb{z}+\textttb{<a>}A_{\text{XML}}\textttb{</a>})^*
\end{align*}
\item Oracle $\mathcal{O}_{\text{XML}}(\alpha)=\mathbb{I}[\alpha\in\mathcal{L}(C_{\text{XML}})]$
\item Seed inputs $E_{\text{XML}}=\{\alpha_{\text{XML}}\}$, where $\alpha_{\text{XML}}=\texttta{<a>hi</a>}$
\end{itemize}
\caption{A context-free language $\mathcal{L}(C_{\text{XML}})$ of XML-like strings, along with an oracle $\mathcal{O}_{\text{XML}}$ for this language and a seed input $\alpha_{\text{XML}}$.}
\label{fig:xml}
\end{figure}

Ideally, we would learn $L_*$ exactly, i.e., $\hat{L}=L_*$, but it is impossible to guarantee exact learning~\cite{gold1967language}. Instead, we want $\hat{L}$ to be a good approximation of $L_*$. To measure the approximation quality, we require probability distributions over $L_*$ and $\hat{L}$. In Section~\ref{sec:sample}, we define the distributions we use in detail. Briefly, we convert the context-free grammar into a \emph{probabilistic context-free grammar}, and use the distribution induced by sampling strings in this probabilistic grammar. Then, we measure the quality of $\hat{L}$ as follows:
\begin{definition}
\label{def:precisionrecall}
\rm
Let $\mathcal{P}_{L_*}$ and $\mathcal{P}_{\hat{L}}$ be probability distributions over $L_*$ and $\hat{L}$, respectively. The \emph{precision} of $\hat{L}$ is $\text{Pr}_{\alpha\sim\mathcal{P}_{\hat{L}}}[\alpha\in L_*]$ and the \emph{recall} of $\hat{L}$ is $\text{Pr}_{\alpha\sim\mathcal{P}_{L_*}}[\alpha\in\hat{L}]$ (here, $\alpha\sim\mathcal{P}$ denotes a random sample from $\mathcal{P}$).
\end{definition}
For high precision, a randomly sampled string $\alpha\sim\mathcal{P}_{\hat{L}}$ must be valid with high probability, i.e., $\alpha\in L_*$. For high recall, $\hat{L}$ must contain a randomly sampled valid string $\alpha\sim\mathcal{P}_{L_*}$ with high probability. Both are desirable: $\hat{L}=\{\alpha_{\text{in}}\}$ has perfect precision but typically low recall, whereas $\hat{L}=\Sigma^*$ has perfect recall but typically low precision. Finally, while the synthesized language $\hat{L}$ is context-free, it is often possible for $\hat{L}$ to approximate $L_*$ with high precision and recall even if $L_*$ is not context-free (e.g., $L_*$ is context-sensitive).

\section{Overview}
\label{sec:algo}

\begin{algorithm}[t]
\footnotesize
\begin{algorithmic}
\Procedure{LearnLanguage}{$\alpha_{\text{in}},\mathcal{O}$}
\State $\hat{L}_{\text{current}}\gets\{\alpha_{\text{in}}\}$
\While{{\bf true}}
\State $M\gets$~\Call{ConstructCandidates}{$\hat{L}_{\text{current}}$}
\State $\tilde{L}_{\text{chosen}}\gets\emptyset$
\ForAll{$\tilde{L}\in M$}
\State $S\gets$~\Call{ConstructChecks}{$\hat{L}_{\text{current}},\tilde{L}$}
\If{\Call{CheckCandidate}{$S,\mathcal{O}$}}
\State $\tilde{L}_{\text{chosen}}\gets\tilde{L}$
\State {\bf break}
\EndIf
\EndFor
\If{$\tilde{L}_{\text{chosen}}=\emptyset$}
\State\Return $\hat{L}_{\text{current}}$
\EndIf
\State $\hat{L}_{\text{current}}\gets\tilde{L}_{\text{chosen}}$
\EndWhile
\EndProcedure
\Procedure{CheckCandidate}{$S,\mathcal{O}$}
\ForAll{$\alpha\in S$}
\If{$\mathcal{O}(\alpha)=0$}
\State\Return{\bf false}
\EndIf
\EndFor
\State\Return{\bf true}
\EndProcedure
\end{algorithmic}
\caption{\small Our grammar synthesis algorithm. Given seed input $\alpha_{\text{in}}\in L_*$ and oracle $\mathcal{O}$ for $L_*$, it returns an approximation of $L_*$.}
\label{alg:infer}
\end{algorithm}

\begin{figure*}
\footnotesize
\centering
\begin{tabular}{llll}
\hline
\multicolumn{1}{c}{{\bf Step}} & \multicolumn{1}{c}{{\bf Language}} & \multicolumn{1}{c}{{\bf Candidates}} & \multicolumn{1}{c}{{\bf Checks}} \\
\hline
{\bf R1}
&
$[\textttb{<a>hi</a>}]_{\text{rep}}$
&
$\begin{array}{l}
\star~([\textttb{<a>hi</a>}]_{\text{alt}})^* \\
{\color{white}\star}~([\textttb{<a>hi</a}]_{\text{alt}})^*[\textttb{>}]_{\text{rep}} \\
{\color{white}\star}~... \\
{\color{white}\star}~\textttb{<a>}([\textttb{hi}]_{\text{alt}})^*[\textttb{</a>}]_{\text{rep}} \\
{\color{white}\star}~...
\end{array}$
&
$\begin{array}{l}
\{\epsilon~\cmark,~\texttta{<a>hi</a><a>hi</a>}~\cmark\} \\
\{\texttta{<a>hi</a}~\xmark,~\texttta{<a>hi</a<a>hi</a>}~\xmark\} \\
... \\
\{\texttta{<a></a>}~\cmark,~\texttta{<a>hihi</a>}~\cmark\} \\
...
\end{array}$ \\
\hline
R2
&
$([\textttb{<a>hi</a>}]_{\text{alt}})^*$
&
$\begin{array}{l}
{\color{white}\star}~([\textttb{<}]_{\text{rep}}+[\textttb{a>hi</a>}]_{\text{alt}})^* \\
{\color{white}\star}~... \\
\star~([\textttb{<a>hi</a>}]_{\text{rep}})^*
\end{array}$
&
$\begin{array}{l}
\{\texttta{<}~\xmark,~\texttta{a>hi</a>}~\xmark\} \\
... \\
\emptyset
\end{array}$ \\
\hline
{\bf R3}
&
$([\textttb{<a>hi</a>}]_{\text{rep}})^*$
&
$\begin{array}{l}
{\color{white}\star}~(([\textttb{<a>hi</a}]_{\text{alt}})^*[\textttb{>}]_{\text{rep}})^* \\
{\color{white}\star}~... \\
\star~(\textttb{<a>}([\textttb{hi}]_{\text{alt}})^*[\textttb{</a>}]_{\text{rep}})^* \\
{\color{white}\star}~...
\end{array}$
&
$\begin{array}{l}
\{\texttta{<a>hi</a}~\xmark,~\texttta{<a>hi</a<a>hi</a>}~\xmark\} \\
... \\
\{\texttta{<a></a>}~\cmark,~\texttta{<a>hihi</a>}~\cmark\} \\
...
\end{array}$ \\
\hline
R4
&
$(\textttb{<a>}([\textttb{hi}]_{\text{alt}})^*[\textttb{</a>}]_{\text{rep}})^*$
&
$\begin{array}{l}
{\color{white}\star}~(\textttb{<a>}([\textttb{hi}]_{\text{alt}})^*([\textttb{</a>}]_{\text{alt}})^*)^* \\
{\color{white}\star}~... \\
{\color{white}\star}~(\textttb{<a>}([\textttb{hi}]_{\text{alt}})^*\textttb{</a}([\textttb{>}]_{\text{alt}})^*)^* \\
\star~(\textttb{<a>}([\textttb{hi}]_{\text{alt}})^*\textttb{</a>})^*
\end{array}$
&
$\begin{array}{l}
\{\texttta{<a>hi}~\xmark,~\texttta{<a>hi</a></a>}~\xmark\} \\
... \\
\{\texttta{<a>hi</a}~\xmark,~\texttta{<a>hi</a>>}~\xmark\} \\
\emptyset
\end{array}$ \\
\hline
{\bf R5}
&
$(\textttb{<a>}([\textttb{hi}]_{\text{alt}})^*\textttb{</a>})^*$
&
$\begin{array}{l}
\star~(\textttb{<a>}([\textttb{h}]_{\text{rep}}+[\textttb{i}]_{\text{alt}})^*\textttb{</a>})^* \\
{\color{white}\star}~(\textttb{<a>}([\textttb{hi}]_{\text{rep}})^*\textttb{</a>})^* \\
\end{array}$
&
$\begin{array}{l}
\{\texttta{<a>h</a>}~\cmark,~\texttta{<a>i</a>}~\cmark\} \\
\emptyset
\end{array}$ \\
\hline
R6
&
$(\textttb{<a>}([\textttb{h}]_{\text{rep}}+[\textttb{i}]_{\text{alt}})^*\textttb{</a>})^*$
&
$\begin{array}{l}
\star~(\textttb{<a>}([\textttb{h}]_{\text{rep}}+[\textttb{i}]_{\text{rep}})^*\textttb{</a>})^*
\end{array}$
&
$\begin{array}{l}
\emptyset
\end{array}$ \\
\hline
R7
&
$(\textttb{<a>}([\textttb{h}]_{\text{rep}}+[\textttb{i}]_{\text{rep}})^*\textttb{</a>})^*$
&
$\begin{array}{l}
\star~(\textttb{<a>}([\textttb{h}]_{\text{rep}}+\textttb{i})^*\textttb{</a>})^*
\end{array}$
&
$\begin{array}{l}
\emptyset
\end{array}$ \\
\hline
R8
&
$(\textttb{<a>}([\textttb{h}]_{\text{rep}}+\textttb{i})^*\textttb{</a>})^*$
&
$\begin{array}{l}
\star~(\textttb{<a>}(\textttb{h}+\textttb{i})^*\textttb{</a>})^*
\end{array}$
&
$\begin{array}{l}
\emptyset
\end{array}$ \\
\hline
R9
&
$(\textttb{<a>}(\textttb{h}+\textttb{i})^*\textttb{</a>})^*$
&
\multicolumn{1}{c}{--}
&
\multicolumn{1}{c}{--} \\
\hline
C1
&
\hspace{-3pt}
$\left(
\hspace{-5pt}
\begin{array}{l}
A_{\text{R1}}'\to(\textttb{<a>}A_{\text{R3}}'\textttb{</a>})^* \\
A_{\text{R3}}'\to(\textttb{h}+\textttb{i})^* \\
\end{array}
\hspace{-5pt},\hspace{3pt}
\{(A_{\text{R1}}',A_{\text{R3}}')\}
\right)$
&
$\begin{array}{l}
\star~
\hspace{-3pt}
\left(
\hspace{-5pt}
\begin{array}{l}
A\to(\textttb{<a>}A\textttb{</a>})^* \\
A\to(\textttb{h}+\textttb{i})^* \\
\end{array}
\hspace{-5pt},\hspace{3pt}
\emptyset
\right) \\
{\color{white}\star}~
\hspace{-3pt}
\left(
\hspace{-5pt}
\begin{array}{l}
A_{\text{R1}}'\to(\textttb{<a>}A_{\text{R3}}'\textttb{</a>})^* \\
A_{\text{R3}}'\to(\textttb{h}+\textttb{i})^* \\
\end{array}
\hspace{-5pt},\hspace{3pt}
\emptyset
\right) \\
\end{array}$
&
$\begin{array}{l}
\{\texttta{hihi}~\cmark,~\texttta{<a><a>hi</a><a>hi</a></a>}~\cmark\} \\\\
\emptyset
\end{array}$ \\
\hline
{\bf C2}
&
\hspace{-3pt}
$\left(
\hspace{-5pt}
\begin{array}{l}
A\to(\textttb{<a>}A\textttb{</a>})^* \\
A\to(\textttb{h}+\textttb{i})^*
\end{array}
\hspace{-5pt},\hspace{3pt}
\emptyset
\right)$
&
\multicolumn{1}{c}{--}
&
\multicolumn{1}{c}{--} \\
\hline
\end{tabular}
\caption{The generalization steps taken by our algorithm given seed input $\alpha_{\text{XML}}$ and oracle $\mathcal{O}_{\text{XML}}$. The initial language $\{\alpha_{\text{XML}}\}$ is generalized to a regular expression in steps R1-R9. The resulting regular expression is translated to a context-free grammar, which is further generalized in steps C1-C2. The candidates at each step are shown in order of preference, with the most preferable on top (ellipses indicate omitted candidates). Checks for each candidate are shown; a green check mark \cmark~ indicates that the check passes and a red cross \xmark~ indicates that it fails. A star $\star$ is shown next to the selected candidate.}
\label{fig:example}
\end{figure*}

In this section, we give an overview of our grammar synthesis algorithm (summarized in Algorithm~\ref{alg:infer}). We consider the case where $E_{\text{in}}$ consists of a single seed input $\alpha_{\text{in}}\in L_*$; an extension to multiple seed inputs is given in Section~\ref{sec:multipleinputs}. Our algorithm starts with the language $\hat{L}_1=\{\alpha_{\text{in}}\}$ containing only the seed input, and constructs a series of languages
\begin{align*}
\{\alpha_{\text{in}}\}=\hat{L}_1\Rightarrow\hat{L}_2\Rightarrow...,
\end{align*}
where $\hat{L}_{i+1}$ results from applying a \emph{generalization step} to $\hat{L}_i$. On one hand, we want the languages to become successively larger (i.e., $\hat{L}_i\subseteq\hat{L}_{i+1}$); on the other hand, we want to avoid overgeneralizing (ideally, the newly added strings $\hat{L}_{i+1}\setminus\hat{L}_i$ should be contained in $L_*$). Our framework returns the current language $\hat{L}_i$ if it is unable to generalize $\hat{L}_i$ in any way. Figure~\ref{fig:example} shows the series of languages constructed by our algorithm for the example in Figure~\ref{fig:xml}. Steps R1-R9 (detailed in Section~\ref{sec:phaseone}) generalize the initial language $\hat{L}_1=\{\alpha_{\text{XML}}\}$ by adding repetitions and alternations. Steps C1-C2 (detailed in Section~\ref{sec:phasetwo}) add recursive productions.

We now describe generalization steps at a high level.

\paragraph{Candidates.}
The $i$th generalization step first constructs \emph{candidate} languages $\tilde{L}_1,...,\tilde{L}_n$, with the goal of choosing $\hat{L}_{i+1}$ to be the candidate that increases recall the most without sacrificing precision. To ensure candidates can only increase recall, we consider \emph{monotone} candidates $\tilde{L}\supseteq\hat{L}_i$. Furthermore, the candidates are ranked from most preferable ($\tilde{L}_1$) to least preferable ($\tilde{L}_n$). Figure~\ref{fig:example} shows the candidates considered for our running example. They are listed in order of preference, with the top candidate being the most preferred. In steps R1-R9, the candidates add a single repetition or alternation to the current regular expression; in steps C1-C2, the candidates try to equate nonterminals in the current context-free grammar.

\paragraph{Checks.}
To ensure high precision, we want to avoid overgeneralizing. Ideally, we want to select a candidate that is \emph{precision-preserving}, i.e., $\tilde{L}\setminus\hat{L}_i\subseteq L_*$. In other words, all strings added to the candidate $\tilde{L}$ (compared to the current language $\hat{L}_i$) are contained in the target language $L_*$. However, we only have access to a membership oracle for $L_*$, so it is typically impossible to prove that a given candidate $\tilde{L}$ is precision-preserving---we would have to check $\mathcal{O}(\alpha)=1$ for every $\alpha\in\tilde{L}\setminus\hat{L}_i$, but this set is often infinite.

Instead, we carefully choose a finite number of heuristic \emph{checks} $S\subseteq\tilde{L}\setminus\hat{L}_i$. Then, our algorithm rejects $\tilde{L}$ if $\mathcal{O}(\alpha)=0$ for any $\alpha\in S$. Alternatively, if all checks pass (i.e., $\mathcal{O}(\alpha)=1$), then $\tilde{L}$ is \emph{potentially precision-preserving}. Since the candidates are ranked in order of preference, we choose the first potentially precision-preserving candidate. Figure~\ref{fig:example} shows examples of checks our algorithm constructs.

\section{Phase One: Regular Expression Synthesis}
\label{sec:phaseone}

We describe the first phase of generalization steps, which generalize the seed input into a regular expression.

\subsection{Candidates}
\label{sec:phaseonecandidate}

In phase one, the current language is represented by a regular expression annotated with extra data: substrings of terminals $\alpha=\sigma_1...\sigma_k$ may be enclosed in square brackets, i.e., $[\blue{\alpha}]_{\tau}$, where $\tau\in\{\text{rep},~\text{alt}\}$. These annotations indicate that the bracketed substring in the current regular expression can be generalized by adding either a repetition (if $\tau=\text{rep}$) or an alternation (if $\tau=\text{alt}$). The seed input $\alpha_{\text{in}}$ is automatically annotated as $[\blue{\alpha_{\text{in}}}]_{\text{rep}}$. Then, each generalization step selects a single bracketed substring $[\blue{\alpha}]_{\tau}$ and generates candidates based on \emph{decompositions} of $\alpha$ (i.e., an expression of $\alpha$ as a sequence of substrings $\alpha=\alpha_1...\alpha_k$):
\begin{itemize}
\item {\bf Repetitions:} If generalizing $P[\blue{\alpha}]_{\text{rep}}Q$, for each decomposition $\alpha=\alpha_1\alpha_2\alpha_3$ such that $\alpha_2\not=\epsilon$, generate
\begin{align*}
P\blue{\alpha_1}([\blue{\alpha_2}]_{\text{alt}})^*[\blue{\alpha_3}]_{\text{rep}}Q.
\end{align*}
\item {\bf Alternations:} If generalizing $P[\blue{\alpha}]_{\text{alt}}Q$, for each decomposition $\alpha=\alpha_1\alpha_2$, where $\alpha_1\not=\epsilon$ and $\alpha_2\not=\epsilon$, generate
\begin{align*}
P([\blue{\alpha_1}]_{\text{rep}}+[\blue{\alpha_2}]_{\text{alt}})Q.
\end{align*}
\end{itemize}
In both cases, the candidate $P\blue{\alpha}Q$ is also generated. For example, in Figure~\ref{fig:example}, step R1 selects $[\textttb{<a>hi</a>}]_{\text{rep}}$ and applies the repetition rule.

The candidates are monotonic (proven in Appendix~\ref{sec:phaseonecandidateproof}):
\begin{proposition}
\label{prop:phaseonecandidate}
\rm
Each candidate constructed in phase one of our algorithm is monotone.
\end{proposition}

We briefly describe the intuition behind these rules. In particular, we define a \emph{meta-grammar}\footnote{We use the term \emph{meta-grammar} to distinguish $\mathcal{C}_{\text{regex}}$ from the context-free grammars we synthesize.} $\mathcal{C}_{\text{regex}}$, which is a context-free grammar whose members $R\in\mathcal{L}(\mathcal{C}_{\text{regex}})$ are regular expressions. The terminals of $\mathcal{C}_{\text{regex}}$ are $\Sigma_{\text{regex}}=\Sigma\cup\{+,*\}$, where $+$ denotes alternations and $*$ denotes repetitions. The nonterminals are $\mathcal{V}_{\text{regex}}=\{T_{\text{rep}},T_{\text{alt}}\}$, where $T_{\text{rep}}$ corresponds to repetitions (and is also the start symbol) and $T_{\text{alt}}$ corresponds to alternations. The productions are
\begin{align*}
T_{\text{rep}}&::=\beta\mid T_{\text{alt}}^*\mid\beta T_{\text{alt}}^*\mid T_{\text{alt}}^*T_{\text{rep}}\mid\beta T_{\text{alt}}^*T_{\text{rep}} \\
T_{\text{alt}}&::=T_{\text{rep}}\mid T_{\text{rep}}+T_{\text{alt}}
\end{align*}
where $\beta\in\Sigma^*-\{\epsilon\}$ ranges over nonempty substrings of $\alpha_{\text{in}}$.

Consider the series of regular expressions $R_1\Rightarrow...\Rightarrow R_n$ in phase one. For each regular expression, we can replace each bracketed substring $[\blue{\alpha}]_{\tau}$ with the nonterminal $T_{\tau}$. Doing so produces a derivation in $\mathcal{C}_{\text{regex}}$, for example, steps R1-R9 in Figure~\ref{fig:example} correspond to the derivation:
\begin{small}
\begin{alignat*}{2}
& [\textttb{<a>hi</a>}]_{\text{rep}}\hspace{0.1in}  && T_{\text{rep}} \\
& \Rightarrow ([\textttb{<a>hi</a>}]_{\text{alt}})^*\hspace{0.1in}  && \Rightarrow T_{\text{alt}}^* \\
& \Rightarrow ([\textttb{<a>hi</a>}]_{\text{rep}})^*\hspace{0.1in}  && \Rightarrow T_{\text{rep}}^* \\
& \Rightarrow (\textttb{<a>}([\textttb{hi}]_{\text{alt}})^*[\textttb{</a>}]_{\text{rep}})^*\hspace{0.1in} && \Rightarrow (\textttb{<a>}T_{\text{alt}}^*T_{\text{rep}})^* \\
& \Rightarrow (\textttb{<a>}([\textttb{hi}]_{\text{alt}})^*\textttb{</a>})^*\hspace{0.1in} && \Rightarrow (\textttb{<a>}T_{\text{alt}}^*\textttb{</a>})^* \\
& \Rightarrow (\textttb{<a>}([\textttb{h}]_{\text{rep}}+[\textttb{i}]_{\text{alt}})^*\textttb{</a>})^*\hspace{0.1in} && \Rightarrow (\textttb{<a>}(T_{\text{rep}}+T_{\text{alt}})^*\textttb{</a>})^* \\
& \Rightarrow (\textttb{<a>}([\textttb{h}]_{\text{rep}}+[\textttb{i}]_{\text{rep}})^*\textttb{</a>})^*\hspace{0.1in} && \Rightarrow (\textttb{<a>}(T_{\text{rep}}+T_{\text{rep}})^*\textttb{</a>})^* \\
& \Rightarrow (\textttb{<a>}([\textttb{h}]_{\text{rep}}+\textttb{i})^*\textttb{</a>})^*\hspace{0.1in} && \Rightarrow (\textttb{<a>}(T_{\text{rep}}+\textttb{i})^*\textttb{</a>})^* \\
%& \Rightarrow ... && \Rightarrow ... \\
& \Rightarrow (\textttb{<a>}(\textttb{h}+\textttb{i})^*\textttb{</a>})^*\hspace{0.1in} && \Rightarrow (\textttb{<a>}(\textttb{h}+\textttb{i})^*\textttb{</a>})^*
\end{alignat*}
\end{small}
In fact, this correspondence goes backwards as well:
\begin{proposition}
\label{prop:correspondence}
\rm
For any derivation $T_{\text{rep}}\xRightarrow{*}R$ in $\mathcal{C}_{\text{regex}}$ (where $R\in\mathcal{L}(\mathcal{C}_{\text{regex}})$), there exists $\alpha_{\text{in}}\in\mathcal{L}(R)$ such that $R$ can be derived from $\alpha_{\text{in}}$ via a series of generalization steps
\begin{align*}
\{\alpha_{\text{in}}\}=R_1\Rightarrow...\Rightarrow R_n=R
\end{align*}
\end{proposition}
We give a proof in Appendix~\ref{sec:correspondenceproof}. Furthermore, $\mathcal{L}(\mathcal{C}_{\text{regex}})$ almost contains every regular expression:
\begin{proposition}
\label{prop:metagrammar}
\rm
For any regular language $L_*$, there exist $R_1,...,R_m\in\mathcal{L}(\mathcal{C}_{\text{regex}})$ such that $L_*=\mathcal{L}(R_1+...+R_m)$.
\end{proposition}
We give a proof in Appendix~\ref{sec:metagrammarproof}. In other words, phase one can synthesize almost any regular language $L_*$, assuming the ``right'' sequence of generalization steps is taken. Our extension to multiple inputs in Section~\ref{sec:multipleinputs} extends this result to any regular language. However, the space of all regular expressions is too large to search exhaustively. We sacrifice completeness for efficiency---our algorithm greedily chooses the first candidate according to the candidate ordering described in Section~\ref{sec:phaseoneorder}.

The productions in $\mathcal{C}_{\text{regex}}$ are unambiguous, so each regular expression $R\in\mathcal{L}(\mathcal{C}_{\text{regex}})$ has a single valid parse tree. This disambiguation allows our algorithm to avoid considering candidate regular expressions multiple times.

\subsection{Candidate Ordering}
\label{sec:phaseoneorder}

The candidate ordering is a heuristic designed to maximize the generality of the regular expression synthesized at the end of phase one. We use the following ordering for candidates constructed by phase one generalization steps:
\begin{itemize}
\item {\bf Repetitions:} If generalizing $P[\blue{\alpha}]_{\text{rep}}Q$, among
\begin{align*}
P\blue{\alpha_1}([\blue{\alpha_2}]_{\text{alt}})^*[\blue{\alpha_3}]_{\text{rep}}Q,
\end{align*}
we first prioritize shorter $\alpha_1$, since $\alpha_1$ is not further generalized. Second, we prioritize longer $\alpha_2$---for example, in step R3 of Figure~\ref{fig:example}, if we instead chose candidate $\textttb{<a>}([\textttb{h}]_{\text{alt}})^*[\textttb{i</a>}]_{\text{rep}}$, then we would synthesize $(\textttb{<a>}\textttb{h}^*\textttb{i}^*\textttb{</a>})^*$, which is less general than step R9.
\item {\bf Alternations:} If generalizing $P[\blue{\alpha}]_{\text{alt}}Q$, among
\begin{align*}
P([\blue{\alpha_1}]_{\text{rep}}+[\blue{\alpha_2}]_{\text{alt}})Q,
\end{align*}
we prioritize shorter $\alpha_1$---for example, in step R5 of Figure~\ref{fig:example}, if we instead chose candidate $(\textttb{<a>}([\textttb{hi}]_{\text{rep}})^*\textttb{</a>})^*$, then step R6 would instead be $(\textttb{<a>}([\textttb{hi}]_{\text{rep}})^*\textttb{</a>})^*$, which is less general than the one we obtain.
\end{itemize}
In either case, the final candidate $P\blue{\alpha}Q$ is ranked last. Note that candidate repetitions and candidate alternations can be ordered independently---each generalization step considers only repetitions (if the chosen bracketed string has form $[\blue{\alpha}]_{\text{rep}}$) or only alternations (if it has form $[\blue{\alpha}]_{\text{alt}}$).

\subsection{Check Construction}
\label{sec:phaseonecheck}

We describe how phase one of our algorithm constructs checks $S\subseteq\tilde{L}\setminus\hat{L}_i$. Each check $\alpha\in S$ has form $\alpha=\gamma\rho\delta$, where $\rho$ is a \emph{residual} capturing the portion of $\tilde{L}$ that is generalized compared to $\hat{L}_i$, and $(\gamma,\delta)$ is a \emph{context} capturing the portion of $\tilde{L}$ which is in common with $\hat{L}_i$. More precisely, suppose the current language is $P[\blue{\alpha}]_{\tau}Q$, where $[\blue{\alpha}]_{\tau}$ is chosen to be generalized, and the candidate language is $PR_{\alpha}Q$, i.e., $\alpha$ is generalized to $R_{\alpha}$. Then, a residual $\rho\in\mathcal{L}(R_{\alpha})\setminus\{\alpha\}$ captures how $R_{\alpha}$ is generalized compared to the substring $\alpha$, and a context $(\gamma,\delta)$ captures the semantics of the expressions $(P,Q)$.

We may want to choose $\gamma\in\mathcal{L}(P)$ and $\delta\in\mathcal{L}(Q)$. However, $P$ and $Q$ may not be regular expressions. For example, on step R5 in Figure~\ref{fig:example}, $P=$~``$(\textttb{<a>}$'', $\alpha=$~``$\textttb{hi}$'', and $Q=$~``$\textttb{</a>})^*$'' (the expressions are quoted to emphasize the placement of parentheses). Instead, $P$ and $Q$ form a regular expression when sequenced together, possibly with a string $\alpha'$ in between, i.e., $P\blue{\alpha'}Q$. We want contexts $(\gamma,\delta)$ such that
\begin{align}
\label{eqn:context}
\gamma\alpha'\delta\in\mathcal{L}(P\blue{\alpha'}Q)\hspace{0.2in}(\forall\alpha'\in\Sigma^*).
\end{align}
Then, the constructed check $\alpha=\gamma\rho\delta$ satisfies
\begin{align*}
\gamma\rho\delta\in\mathcal{L}(P\blue{\rho}Q)\subseteq\mathcal{L}(PR_{\alpha}Q),
\end{align*}
where the first inclusion follows from (\ref{eqn:context}) and the second inclusion follows since $\rho\in\mathcal{L}(R_{\alpha})$. We discard $\alpha$ such that $\alpha\in\mathcal{L}(\hat{L}_i)$ to obtain valid checks $\alpha\in\tilde{L}\setminus\hat{L}_i$.

Next, we explain the construction of residuals and contexts. Our algorithm generates residuals as follows:
\begin{itemize}
\item {\bf Repetitions:} For current language $P[\blue{\alpha}]_{\text{rep}}Q$ and candidate $P\blue{\alpha_1}([\blue{\alpha_2}]_{\text{alt}})^*[\blue{\alpha_3}]_{\text{rep}}Q$, generate residuals $\alpha_1\alpha_3$ and $\alpha_1\alpha_2\alpha_2\alpha_3$.
\item {\bf Alternations:} For current language $P[\blue{\alpha}]_{\text{alt}}Q$ and candidate $P(\blue{\alpha_1}+\blue{\alpha_2})Q$, generate residuals $\alpha_1$ and $\alpha_2$.
\end{itemize}

Next, our algorithm associates a context $(\gamma,\delta)$ with each bracketed string $[\blue{\alpha}]_{\tau}$. The context for the initial bracketed string $[\blue{\alpha_{\text{in}}}]_{\text{rep}}$ is $(\epsilon,\epsilon)$. After each generalization step, contexts for new bracketed substrings are generated:
\begin{itemize}
\item {\bf Repetitions:} For current language $P[\blue{\alpha}]_{\text{rep}}Q$, where $[\blue{\alpha}]_{\text{rep}}$ has context $(\gamma,\delta)$, and candidate $P\blue{\alpha_1}([\blue{\alpha_2}]_{\text{alt}})^*[\blue{\alpha_3}]_{\text{rep}}Q$, the context generated for the new bracketed substring $[\blue{\alpha_2}]_{\text{alt}}$ is $(\gamma\alpha_1,\alpha_3\delta)$, and for $[\blue{\alpha_3}]_{\text{rep}}$ is $(\gamma\alpha_1\alpha_2,\delta)$.
\item {\bf Alternations:} For current language $P[\blue{\alpha}]_{\text{alt}}Q$, where $[\blue{\alpha}]_{\text{alt}}$ has context $(\gamma,\delta)$, and candidate $P([\blue{\alpha_1}]_{\text{rep}}+[\blue{\alpha_2}]_{\text{alt}})Q$, the context generated for the new bracketed substring $[\blue{\alpha_1}]_{\text{rep}}$ is $(\gamma,\alpha_2\delta)$, and for $[\blue{\alpha_2}]_{\text{alt}}$ is $(\gamma\alpha_1,\delta)$.
\end{itemize}

For example, on step R3, the context for $[\textttb{<a>hi</a>}]_{\text{rep}}$ is $(\epsilon,\epsilon)$. The residuals for candidate $(([\textttb{<a>hi</a}]_{\text{alt}})^*[\textttb{>}]_{\text{rep}})^*$ are $\texttta{<a>hi</a}$ and $\texttt{<a>hi</a>>}$; since the context is empty, these residuals are also the checks, and they are rejected by the oracle, so the candidate is rejected. On the other hand, the residuals (and checks) for the chosen candidate $(\textttb{<a>}([\textttb{hi}]_{\text{alt}})^*[\textttb{</a>}]_{\text{rep}})^*$ are $\texttta{<a></a>}$ and $\texttta{<a>hihi</a>}$, which are accepted by the oracle. For the new bracketed string $[\textttb{hi}]_{\text{alt}}$, the algorithm constructs the context $(\texttta{<a>},\texttta{</a>})$, and for the new bracketed string $[\textttb{</a>}]_{\text{rep}}$, the algorithm constructs the context $(\texttta{<a>hi},\epsilon)$.

Similarly, on step R5, the context for $[\textttb{hi}]_{\text{alt}}$ is $(\texttta{<a>},\texttta{</a>})$. The residuals constructed for the chosen candidate $(\textttb{<a>}([\textttb{h}]_{\text{rep}}+[\textttb{i}]_{\text{alt}})^*\textttb{</a>})^*$ are $\texttta{h}$ and $\texttta{i}$, so the constructed checks are $\texttta{<a>h</a>}$ and $\texttta{<a>i</a>}$. Our algorithm constructs the context $(\texttta{<a>},\texttta{i</a>})$ for the new bracketed string $[\textttb{h}]_{\text{rep}}$ and the context $(\texttta{<a>h},\texttta{</a>})$ for the new bracketed string $[\textttb{i}]_{\text{alt}}$.

We have the following result:
\begin{proposition}
\label{prop:context}
\rm
The contexts constructed by phase one generalization steps satisfy (\ref{eqn:context}).
\end{proposition}
We give a proof in Appendix~\ref{sec:contextproof}, which ensures that the constructed checks are valid (i.e., belong to $\tilde{L}\setminus\hat{L}_i$).

\subsection{Computational Complexity}
\label{sec:phaseonecomplexity}

Let $n$ be the length of the seed input $\alpha_{\text{in}}$. In phase one, our algorithm considers at most $O(n^2)$ repetition candidates (since each of the $n^2$ substrings of $\alpha_{\text{in}}$ is considered at most once), and $O(n^3)$ alternation candidates (since at most $O(n)$ alternation candidates are considered per discovered repetition). Examining each candidate takes constant time (assuming each query to $\mathcal{O}$ takes constant time), so the complexity of phase one is $O(n^3)$. In our evaluation, we show that our algorithm is quite scalable.

\section{Phase Two: Recursive Properties}
\label{sec:phasetwo}

\begin{figure*}
\centering
\small
\begin{tabular}{lllll}
\hline
\multicolumn{1}{c}{{\bf Step}} & \multicolumn{2}{c}{{\bf Chosen Generalization}} & \multicolumn{1}{c}{{\bf Productions}} & \multicolumn{1}{c}{{\bf Language $\mathcal{L}(\hat{C},A_i)$}} \\
\hline
{\bf R1}
& $[\textttb{<a>hi</a>}]_{\text{rep}}^{\text{R1}}$ & $\Rightarrow([\textttb{<a>hi</a>}]_{\text{alt}}^{\text{R2}})^*$
& $\{A_{\text{R1}}\to A_{\text{R1}}',\hspace{0.05in}A_{\text{R1}}'\to\epsilon+A_{\text{R1}}'A_{\text{R2}}\}$
& $(\textttb{<a>}(\textttb{h}+\textttb{i})^*\textttb{</a>})^*$
\\
R2
& $[\textttb{<a>hi</a>}]_{\text{alt}}^{\text{R2}}$ & $\Rightarrow[\textttb{<a>hi</a>}]_{\text{rep}}^{\text{R3}}$
& $\{A_{\text{R2}}\to A_{\text{R3}}\}$
& $\textttb{<a>}(\textttb{h}+\textttb{i})^*\textttb{</a>}$
\\
{\bf R3}
& $[\textttb{<a>hi</a>}]_{\text{rep}}^{\text{R3}}$ & $\Rightarrow\textttb{<a>}([\textttb{hi}]_{\text{alt}}^{\text{R5}})^*[\textttb{</a>}]_{\text{rep}}^{\text{R4}}$
& $\{A_{\text{R3}}\to\textttb{<a>}A_{\text{R3}}'A_{\text{R4}},\hspace{0.05in}A_{R3}'\to\epsilon+A_{R3}'A_{R5}\}$
& $\textttb{<a>}(\textttb{h}+\textttb{i})^*\textttb{</a>}$
\\
R4
& $[\textttb{</a>}]_{\text{rep}}^{\text{R4}}$ & $\Rightarrow\textttb{</a>}$
& $\{A_{\text{R4}}\to\textttb{</a>}\}$
& $\textttb{<a>}$
\\
{\bf R5}
& $[\textttb{hi}]_{\text{alt}}^{\text{R5}}$ & $\Rightarrow[\textttb{h}]_{\text{rep}}^{\text{R8}}+[\textttb{i}]_{\text{alt}}^{\text{R6}}$
& $\{A_{\text{R5}}\to A_{\text{R8}}+A_{\text{R6}}\}$
& $\textttb{h}+\textttb{i}$
\\
R6
& $[\textttb{i}]_{\text{alt}}^{\text{R6}}$ & $\Rightarrow[\textttb{i}]_{\text{rep}}^{\text{R7}}$
& $\{A_{\text{R6}}\to A_{\text{R7}}\}$
& $\textttb{i}$
\\
R7
& $[\textttb{i}]_{\text{rep}}^{\text{R7}}$ & $\Rightarrow\textttb{i}$
& $\{A_{\text{R7}}\to\textttb{i}\}$
& $\textttb{i}$
\\
R8
& $[\textttb{h}]_{\text{alt}}^{\text{R8}}$ & $\Rightarrow\textttb{h}$
& $\{A_{\text{R8}}\to\textttb{h}\}$
& $\textttb{h}$
\\
R9 & \multicolumn{2}{c}{--} & \multicolumn{1}{c}{--} & \multicolumn{1}{c}{--} \\
\hline
\end{tabular}
\caption{The productions added to $\hat{C}_{\text{XML}}$ corresponding to each generalization step are shown. The derivation shows the bracketed subexpression $[\blue{\alpha}]_{\tau}^i$ (annotated with the step number $i$) selected to be generalized at step $i$, as well as the subexpression to which $[\blue{\alpha}]_{\tau}^i$ is generalized. The language $\mathcal{L}(\hat{C},A_i)$ (i.e., strings derivable from $A_i$) equals the subexpression in $\hat{R}$ that eventually replaces $[\blue{\alpha}]_{\tau}^i$. As before, steps that select a candidate that strictly generalizes the language are bolded (in the first column).}
\label{fig:exampletranslation}
\end{figure*}

The second phase of generalization steps learn recursive properties of program input languages that cannot be represented using regular expressions. Consider the regular expression $(\textttb{<a>}(\textttb{h}+\textttb{i})^*\textttb{</a>})^*$ obtained at the end of phase one in Figure~\ref{fig:example}, which can be written as $\hat{R}_{\text{XML}}=(\textttb{<a>}R_{\text{hi}}\textttb{</a>})^*$, where $R_{\text{hi}}=(\textttb{h}+\textttb{i})^*$. Since every regular language is also context-free, we can begin by translating $\hat{R}_{\text{XML}}$ to the context-free grammar
\begin{align*}
\{A_{\text{XML}}\to(\textttb{<a>}A_{\text{hi}}\textttb{</a>})^*,~A_{\text{hi}}\to(\textttb{h}+\textttb{i})^*\}.
\end{align*}
Then, we can equate the nonterminals $A_{\text{XML}}$ and $A_{\text{hi}}$ to obtain the context-free grammar $\hat{C}_{\text{XML}}$:
\begin{align*}
\{A\to(\textttb{<a>}A\textttb{</a>})^*,~A\to(\textttb{h}+\textttb{i})^*\},
\end{align*}
which does not overgeneralize, since $\mathcal{L}(\hat{C}_{\text{XML}})\subseteq\mathcal{L}(C_{\text{XML}})$. Furthermore, $\mathcal{L}(\hat{C}_{\text{XML}})$ is not regular, as it contains the language of matching tags $\textttb{<a>}$ and $\textttb{</a>}$.

In general, phase two of algorithm first translates the synthesized regular expression $\hat{R}$ into a context-free grammar $\hat{C}$. Then, each generalization step considers equating a pair $(A,B)$ of nonterminals in $\hat{C}$, where $A$ and $B$ correspond to \emph{repetition subexpressions} of $\hat{R}$, which are subexpressions $R$ of $\hat{R}$ of the form $R=R_1^*$. The restriction to equating repetition subexpressions is empirically motivated---in practice, recursive constructs can typically also be repeated, e.g., in matching parentheses grammars, so constraining the search space reduces the potential for imprecision without sacrificing recall. In our example, $A_{\text{XML}}$ corresponds to repetition subexpression $\hat{R}_{\text{XML}}$, and $A_{\text{hi}}$ corresponds to repetition subexpression $R_{\text{hi}}$, so our algorithm considers equating $A_{\text{XML}}$ and $A_{\text{hi}}$.

In the remainder of this section, we first describe how we translate regular expressions to context-free grammars, and then describe phase two candidates and checks.

\subsection{Translating $\hat{R}$ to a Context-Free Grammar}
\label{sec:translation}

Our algorithm translates the regular expression $\hat{R}$ to a context-free grammar $\hat{C}=(V,\Sigma,P,T)$ such that $\mathcal{L}(\hat{R})=\mathcal{L}(\hat{C})$ and subexpressions in $\hat{R}$ correspond to nonterminals in $\hat{C}$. Intuitively, the translation follows the derivation of $\hat{R}$ in the meta-grammar $\mathcal{C}_{\text{regex}}$ (described in Section~\ref{sec:phaseonecandidate}). First, the terminals in $\hat{C}$ are the program input alphabet $\Sigma$. Next, the nonterminals $V$ of $\hat{C}$ correspond to generalization steps, additionally including an auxiliary nonterminal for steps that generalize repetition nodes:
\begin{align*}
V=\{A_i\mid\text{step }i\}\cup\{A_i'\mid\text{step }i\text{ generalizes }P[\blue{\alpha}]_{\text{rep}}Q\}.
\end{align*}
The start symbol is $A_1$. Finally, the productions are generated according to the following rules:
\begin{itemize}
\item {\bf Repetition:} If step $i$ generalizes current language $P[\blue{\alpha}]_{\text{rep}}Q$ to $P\blue{\alpha_1}([\blue{\alpha_2}]_{\text{alt}})^*[\blue{\alpha_3}]_{\text{rep}}Q$, we generate productions
\begin{align*}
A_i\to\blue{\alpha_1}A_i'A_k,\sss A_i'\to\epsilon+A_i' A_j,
\end{align*}
where $j$ is the step that generalizes $[\blue{\alpha_2}]_{\text{alt}}$ and $k$ is the step that generalizes $[\blue{\alpha_3}]_{\text{rep}}$. Intuitively, these productions are equivalent to the ``production'' $A_i\to\blue{\alpha_1}A_j^*A_k$.
\item {\bf Alternation:} If step $i$ generalizes $P[\blue{\alpha}]_{\text{alt}}Q$ to $P([\blue{\alpha_1}]_{\text{rep}}+[\blue{\alpha_2}]_{\text{alt}})Q$, we include production $A_i\to A_j+A_k$, where $j$ is the step that generalizes $[\blue{\alpha_1}]_{\text{rep}}$ and $k$ is the step that generalizes $[\blue{\alpha_2}]_{\text{alt}}$.
\end{itemize}
For example, Figure~\ref{fig:exampletranslation} shows the result of the translation algorithm applied to the generalization steps in the first phase of Figure~\ref{fig:example} to produce a context-free grammar $\hat{C}_{\text{XML}}$ equivalent to $\hat{R}_{\text{XML}}$. Here, steps R1 and R3 handle the semantics of repetitions, step R5 handles the semantics of the alternation, steps R2 and R6 only affect brackets so they are identities, and steps R4, R7, and R8 are constant expressions. Furthermore, $\mathcal{L}(\hat{C},A_i)$ is the language of strings matched by the subexpression that eventually replaces the bracketed substring $[\blue{\alpha}]_{\tau}$ generalized on step $i$; this language is shown in the last column of Figure~\ref{fig:exampletranslation}.

The auxiliary nonterminals $A_i'$ correspond to repetition subexpressions in $\hat{R}$---if step $i$ generalizes $[\blue{\alpha}]_{\text{rep}}$ to $\blue{\alpha_1}([\blue{\alpha_2}]_{\text{alt}})^*[\blue{\alpha_3}]_{\text{rep}}$, then $\mathcal{L}(\hat{C},A_i')=\mathcal{L}(R^*)$, where $R$ is the subexpression to which $[\blue{\alpha_2}]_{\text{alt}}$ is eventually generalized. In our example, $A_{\text{R1}}'$ corresponds to $\hat{R}_{\text{XML}}=(\textttb{<a>}(\textttb{h}+\textttb{i})^*\textttb{</a>})^*$, and $A_{\text{R3}}'$ corresponds to $R_{\text{hi}}=(\textttb{h}+\textttb{i})^*$.

For conciseness, we redefine $\hat{C}_{\text{XML}}$ to be the equivalent context-free grammar with start symbol $A_{\text{R1}}'$ and productions
\begin{align*}
A_{\text{R1}}'\to(\textttb{<a>}A_{\text{R3}}'\textttb{</a>})^*,\sss A_{\text{R3}}'\to(\textttb{h}+\textttb{i})^*
\end{align*}
where the Kleene star implicitly expands to the productions described in the repetition case.

\subsection{Candidates and Ordering}
\label{sec:phasetwocandidate}

The candidates considered in phase two of our algorithm are \emph{merges}, which are (unordered) pairs of nonterminals $(A_i',A_j')$ in $\hat{C}$, where $i$ and $j$ are generalization steps of phase one. Recall that these nonterminals correspond to repetition subexpressions in $\hat{R}$. In particular, associated to $\hat{C}$ is the set $M$ of all such pairs of nonterminals. In Figure~\ref{fig:example}, the regular expression $\hat{R}_{\text{XML}}$ on step R9 is translated into the context-free grammar $\hat{C}_{\text{XML}}$ on step C1, with its corresponding set of merges $M_{\text{XML}}$ containing just $(A_{\text{R1}}',A_{\text{R3}}')$.

Each phase two generalization step selects a pair $(A_i',A_j')\in M$ and considers two candidates (in order of preference):
\begin{itemize}
\item The first candidate $\tilde{C}$ equates $A_i'$ and $A_j'$ by introducing a fresh nonterminal $A$ and replacing all occurrences of $A_i'$ and $A_j'$ in $\hat{C}$ with $A$.
\item The second candidate equals the current language $\hat{C}$.
\end{itemize}
In either case, the selected pair is removed from $M$. The candidates are monotone since equating two nonterminals can only enlarge the generated language.

For example, in step C1 of Figure~\ref{fig:example}, the candidate $(A_{\text{R1}}',A_{\text{R3}}')$ is removed from $M_{\text{XML}}$; the first candidate is constructed by equating $A_{\text{R1}}'$ and $A_{\text{R3}}'$ in $\hat{C}_{\text{XML}}$ to obtain
\begin{align*}
\tilde{C}_{\text{XML}}=\{A\to(\textttb{<a>}A\textttb{</a>})^*,~A\to(\textttb{h}+\textttb{i})^*\},
\end{align*}
where $\mathcal{L}(\tilde{C}_{\text{XML}})$ is not regular. The chosen candidate is $\hat{C}_{\text{XML}}'=\tilde{C}_{\text{XML}}$, since the checks (described in Section~\ref{sec:phasetwocheck}) pass. On step C2, $M$ is empty, so our algorithm returns $\hat{C}_{\text{XML}}'$. In particular, $\hat{C}_{\text{XML}}'$ equals $\mathcal{L}(C_{\text{XML}})$, except the characters $\textttb{a}+...+\textttb{z}$ are restricted to $\textttb{h}+\textttb{i}$. In Section~\ref{sec:charactergen}, we describe an extension that generalizes characters in $\hat{C}_{\text{XML}}'$.

Finally, we formalize the intuition that equating $(A_i',A_j')\in M$ corresponds to merging repetition subexpressions:
\begin{proposition}
\label{prop:merge}
\rm
Let regular expression $\hat{R}$ translate to context-free grammar $\hat{C}$. Suppose that nonterminal $A_i$ in $\hat{C}$ corresponds to repetition subexpression $R$, so $\hat{R}=PRQ$, and $A_j$ to $R'$, so $\hat{R}=P'R'Q'$. Let $\tilde{C}$ be obtained by equating $A_i$ and $A_j$ in $\hat{C}$. Then, $\mathcal{L}(PR'Q)\subseteq\mathcal{L}(\tilde{C})$ (and symmetrically, $\mathcal{L}(P'RQ')\subseteq\mathcal{L}(\tilde{C})$).
\end{proposition}
In other words, equating $(A_i',A_j')\in M$ merges $R$ and $R'$ in $\hat{R}$. We give a proof in Appendix~\ref{sec:propmergeproof}.

\subsection{Check Construction}
\label{sec:phasetwocheck}

Consider the candidate $\tilde{C}$ obtained by merging $(A_i',A_j')\in M$ in the current language $\hat{C}$, where $A_i'$ corresponds to repetition subexpression $R$ and $A_j'$ to $R'$. Suppose that step $i$ generalizes $P[\blue{\alpha}]_{\text{rep}}Q$ to $\blue{\alpha_1}([\blue{\alpha_2}]_{\text{alt}})^*[\blue{\alpha_3}]_{\text{rep}}$, and step $j$ generalizes $[\blue{\alpha'}]_{\text{rep}}$ to $\blue{\alpha_1'}([\blue{\alpha_2'}]_{\text{alt}})^*[\blue{\alpha_3'}]_{\text{rep}}$. Note that $([\blue{\alpha_2}]_{\text{alt}})^*$ is eventually generalized to the repetition subexpression $R$ in $\hat{R}$, and $([\blue{\alpha_2'}]_{\text{alt}})^*$ is eventually generalized to $R'$ in $\hat{R}$.

Our algorithm constructs the check $\gamma\rho'\delta$, where $\rho'=\alpha'\alpha'\in\mathcal{L}(R')$ is a residual for $R'$, and $(\gamma,\delta)$ is the context for $([\blue{\alpha_2}]_{\text{alt}})^*$. This check satisfies
\begin{align*}
\gamma\rho'\delta\in\mathcal{L}(PR'Q)\subseteq\mathcal{L}(\tilde{C}),
\end{align*}
where the first inclusion follows by the property (\ref{eqn:context}) for contexts described in Section~\ref{sec:phaseonecheck}, and the second inclusion follows from Proposition~\ref{prop:merge}. A similar argument to Proposition~\ref{prop:context} shows that this context satisfies property (\ref{eqn:context}).

The check $\gamma\rho'\delta$ tries to ensure that $R'$ can be substituted for $R$ without overgeneralizing, i.e., $\mathcal{L}(PR'Q)\subseteq L_*$. Our algorithm similarly generates a second check trying to ensure that $R$ can be substituted for $R'$, i.e., $\mathcal{L}(P'RQ)\subseteq L_*$.

For example, in Figure~\ref{fig:example}, the context for the repetition subexpression $\hat{R}_{\text{XML}}=(\textttb{<a>}(\textttb{h}+\textttb{i})^*\textttb{</a>})^*$ is $(\epsilon,\epsilon)$, and the residual for $R_{\text{hi}}$ is $\texttta{hihi}$, so the constructed check is \texttta{hihi}. Similarly, the context for $R_{\text{hi}}$ is $(\texttta{<a>},\texttta{</a>})$ and the residual for $\hat{R}_{\text{XML}}$ is \texttta{<a>hi</a><a>hi</a>}, so the constructed check is \texttta{<a><a>hi</a><a>hi</a></a>}.

\subsection{Learning Matching Parentheses Grammars}
\label{sec:phasetwoexpressiveness}

To demonstrate the expressive power of merges, we show that they can represent the following class of generalized matching parentheses grammars:
\begin{definition}
\label{def:matchingparen}
\rm
A \emph{generalized matching parentheses grammar} is a context-free grammar $C=(V,\Sigma,P,S_1)$, with
\begin{align*}
V=\{S_1,...,S_n,R_1,...,R_n,R_1',...,R_n'\}
\end{align*}
and productions
\begin{align*}
S_i\rightarrow(R_i(S_{i_1}+...+S_{i_{k_i}})^*R_i')^*,
\end{align*}
where for $1\le i\le n$, $R_i,R_i'$ are regular expressions over $\Sigma$.
\end{definition}
In other words, $R_i$ and $R_i'$ are pairs of matching parentheses, except that they are allowed to be regular expressions, e.g., XML tags. They may also match the empty string $\epsilon$, e.g., to permit unmatched open parentheses. Then, the valid matched parentheses strings matched by the grammars $S_{i_1},...,S_{i_{k_i}}$ can occur between $R_i$ and $R_i'$. In particular, the XML-like grammar shown in Figure~\ref{fig:xml} is a generalized matching parentheses grammar, where the ``parentheses'' are \texttt{<a>} and \texttt{</a>}. We have the following result:
\begin{proposition}
\label{prop:phasetwoexpressiveness}
\rm
For any generalized matching parentheses grammar $C$, there exists a regular expression $R$ and merges $M$ over $R$ such that letting $C'$ be the grammar obtained by transforming $R$ into a context-free grammar and performing the merges in $M$, we have $\mathcal{L}(C)=\mathcal{L}(C')$.
\end{proposition}
In other words, phase two of our algorithm at least allows us to learn the common class of generalized matching parentheses grammars. We give a proof in Appendix~\ref{sec:phasetwoexpressivenessproof}.

\subsection{Computational Complexity}
\label{sec:phasetwocomplexity}

The complexity of phase two is $O(n^4)$, where $n$ is the length of the seed input $\alpha_{\text{in}}$, since each pair of repetition subexpressions is a merge candidate, and as shown in Section~\ref{sec:phaseonecomplexity}, there are at most $O(n^2)$ repetition candidates. Therefore, the overall complexity is $O(n^4)$.

\section{Extensions}
\label{sec:extensions}

In this section, we discuss two extensions to our algorithm.

\subsection{Multiple Seed Inputs}
\label{sec:multipleinputs}

Given multiple seed inputs $E_{\text{in}}=\{\alpha_1,...,\alpha_n\}$, our algorithm first applies phase one separately to each $\alpha_i$ to synthesize a corresponding regular expression $\hat{R}_i$. Then, it combines these into a single regular expression $\hat{R}=\hat{R}_1+...+\hat{R}_n$ and applies phase two to $\hat{R}$. Repetition subexpressions in different components $\hat{R}_i$ of $\hat{R}$ may be merged. A useful optimization is to construct $\hat{R}$ incrementally---if we have $\alpha_i\in\mathcal{L}(\hat{R}_1+...+\hat{R}_{i-1})$, then $\alpha_i$ can be skipped.

\subsection{Character Generalization}
\label{sec:charactergen}

After phase one, we include a \emph{character generalization} phase that generalizes terminals in the synthesized regular expression $\hat{R}$. At each generalization step, the algorithm selects a terminal string $\alpha=\sigma_1...\sigma_k$ in $\hat{R}$, i.e., $\hat{R}=P\blue{\alpha}Q$, and a terminal $\sigma_i$ in $\alpha$, and a different terminal $\sigma\in\Sigma$ such that $\sigma\not=\sigma_i$, and considers two candidates. First, $\tilde{R}=P\blue{\sigma_1...\sigma_{i-1}}(\blue{\sigma}+\blue{\sigma_i})\blue{\sigma_{i+1}...\sigma_k}Q$ replaces $\sigma_i$ with $(\blue{\sigma_i}+\blue{\sigma})$. Second, the current language $\hat{R}$. Each such generalization is considered exactly once in this phase.

For the first candidate, we construct residual $\rho=\sigma$. Every terminal string $\alpha$ in $\hat{R}$ was added by generalizing $[\blue{\alpha'}_{\text{rep}}]$ to $\blue{\alpha_1}([\blue{\alpha_2}]_{\text{alt}})^*[\blue{\alpha_3}]_{\text{rep}}$, where $\alpha=\alpha_1$. Supposing that the context for $[\blue{\alpha'}_{\text{rep}}]$ is $(\gamma,\delta)$, we construct context $(\gamma\sigma_1...\sigma_{i-1},\sigma_{i+1}...\sigma_k\alpha_3\delta)$. The generated checks are $\gamma\rho\delta$.

For example, in the regular expression $\hat{R}_{\text{XML}}$ output by phase one in Figure~\ref{fig:example}, our algorithm considers generalizing each terminal in \texttta{<a>}, \texttta{h}, \texttta{i}, and \texttta{</a>} to every (different) terminal $\sigma\in\Sigma$. Generalizing \texttta{<} to \texttta{a} is ruled out by the check \texttta{aa>hi</a>}. Alternatively, \texttta{h} is generalized to \texttta{a} since the generated checks \texttta{<a>ai</a>} and \texttta{<a>a</a>} pass. Eventually, $\hat{R}_{\text{XML}}$ generalizes to
\begin{align*}
\hat{R}_{\text{XML}}'=(\textttb{<a>}((\textttb{a}+...+\textttb{z})+(\textttb{a}+...+\textttb{z}))^*\textttb{</a>})^*,
\end{align*}
which phase two generalizes to the grammar $\hat{C}_{\text{XML}}'$:
\begin{align*}
\left\{\begin{array}{l}
A\to(\textttb{<a>}A\textttb{</a>})^*,\\
A\to((\textttb{a}+...+\textttb{z})+(\textttb{a}+...+\textttb{z}))^*
\end{array}\right\}.
\end{align*}
In particular, $\mathcal{L}(\hat{C}_{\text{XML}}')=\mathcal{L}(C_{\text{XML}})$.

\section{Discussion}

\paragraph{Phases of \toolname.}

We have described \toolname as proceeding in three phases, but the distinction is primarily for purposes of clarity. More precisely, the character generalization phase can equivalently be performed at any time. Phase two (the merging phase) depends on phase one to identify candidate repetition subexpressions to merge, but these phases could be interleaved if desired.

\paragraph{Limitations.}

\begin{figure*}
\begin{tabular}{ccc}
\includegraphics[width=0.3\textwidth]{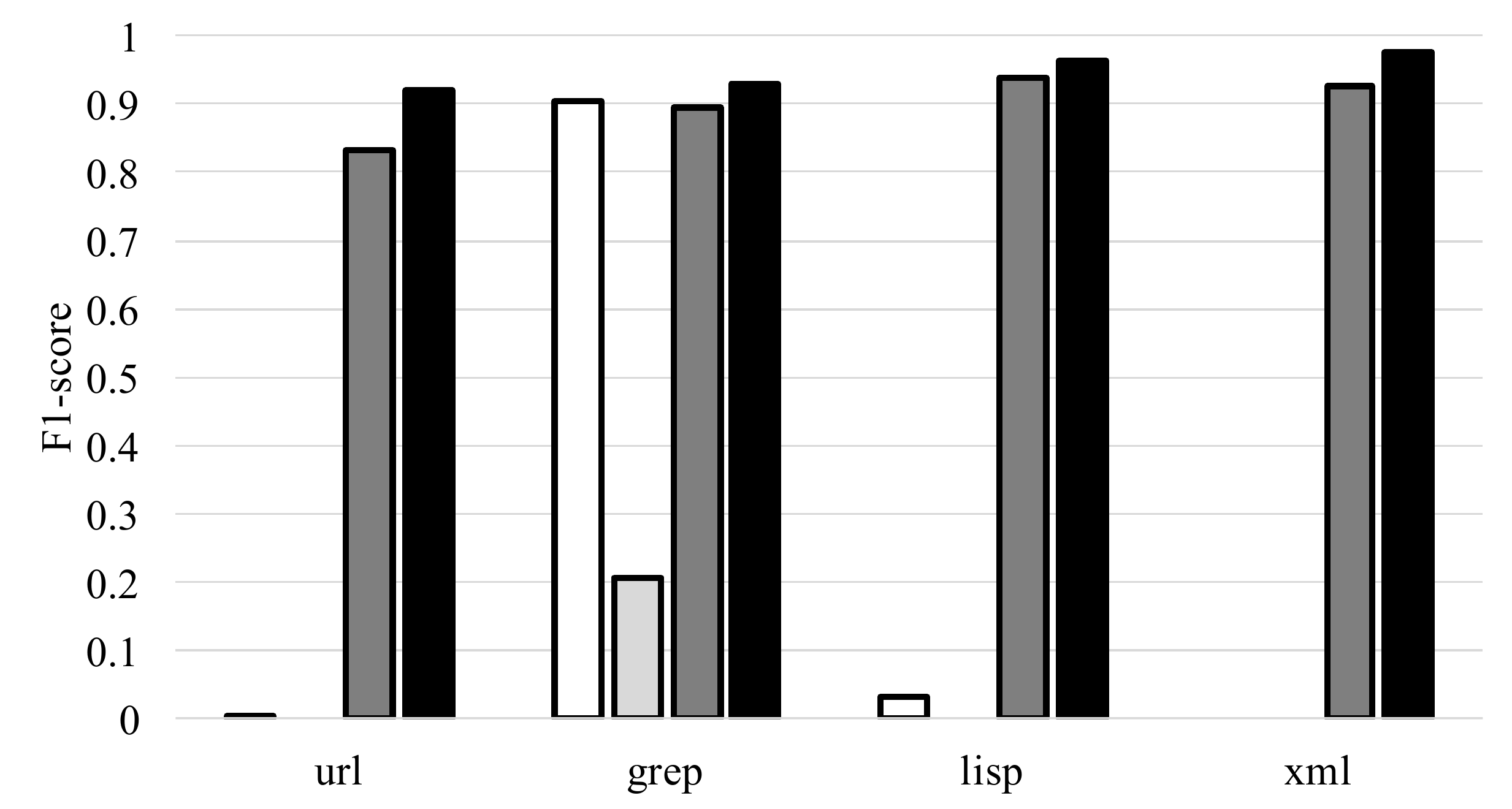}
& \includegraphics[width=0.3\textwidth]{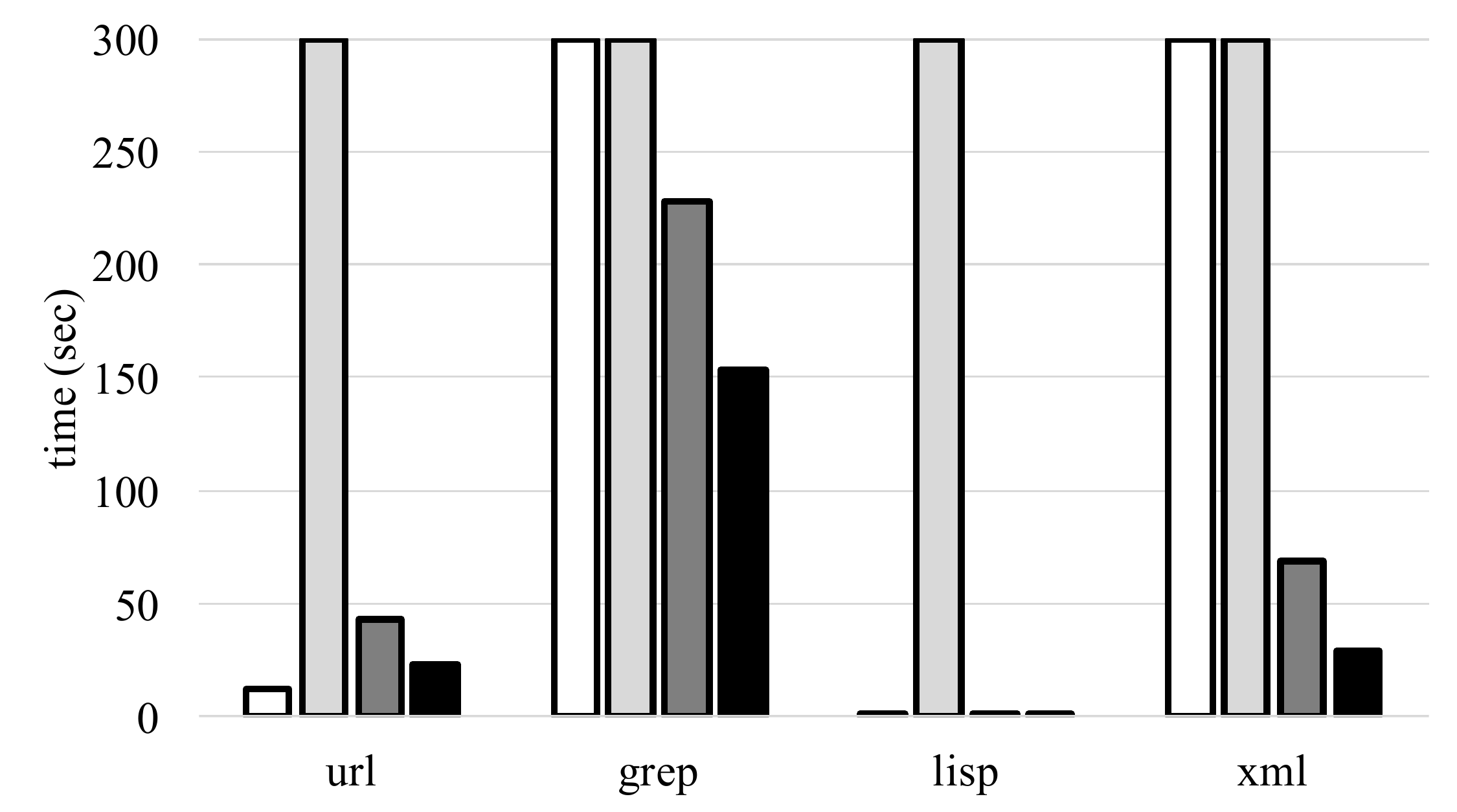}
& \includegraphics[width=0.3\textwidth]{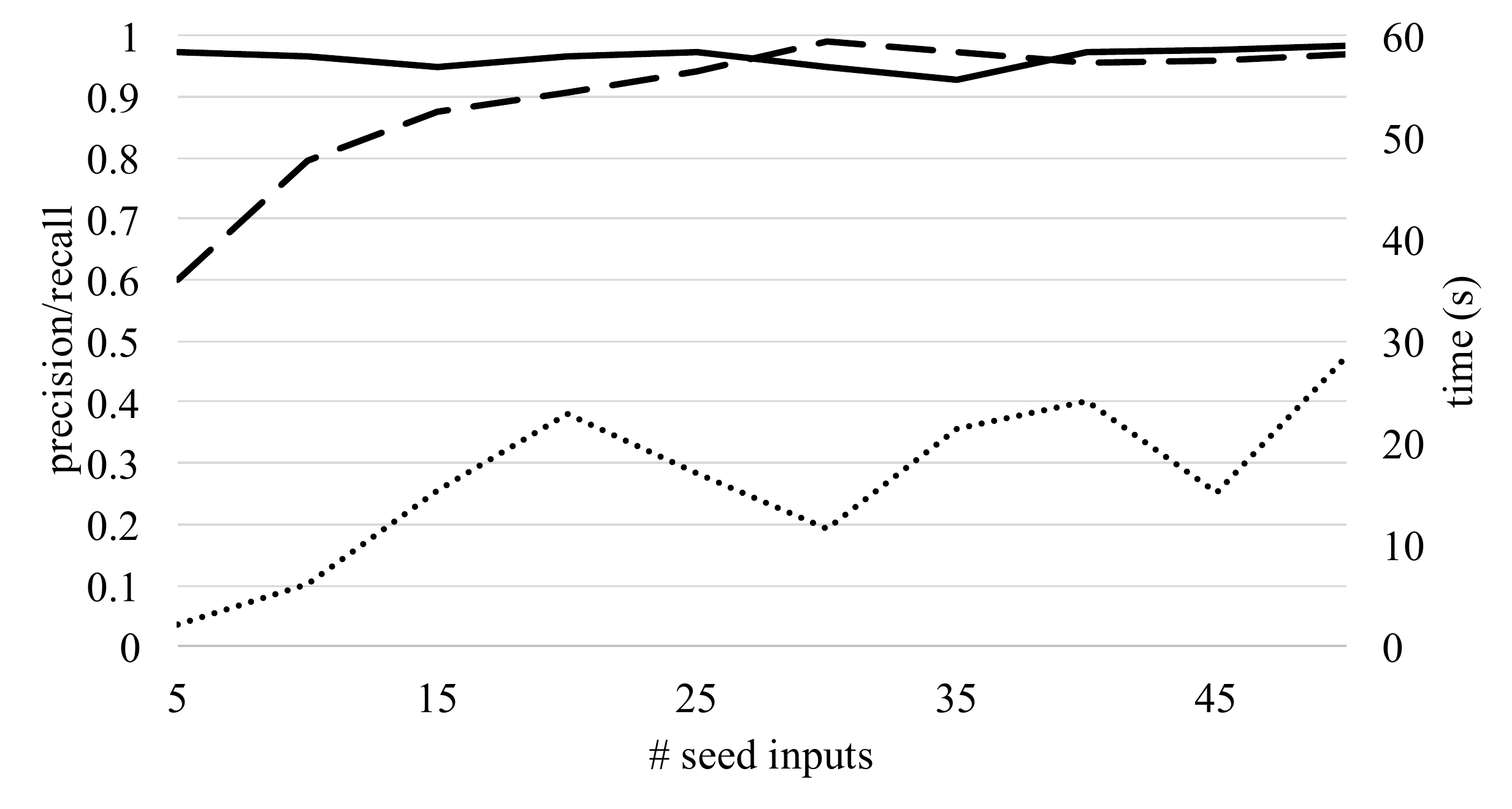} \\
(a) & (b) & (c)
\end{tabular}
\caption{We show (a) the $F_1$ score, and (b) the running time of $L$-Star (white), RPNI (light grey), \toolname omitting phase two (dark grey), and \toolname (black) for each of the four test grammars $C$. The algorithms are trained on 50 random samples from the target language $L_*=\mathcal{L}(C)$. In (c), for the XML grammar, we show how the precision (solid line), recall (dashed line), and running time (dotted line) of \toolname vary with the number of seed inputs $|E_{\text{in}}|$ (between 0 and 50). The $y$-axis for precision and recall is on the left-hand side, whereas the $y$-axis for the running time (in seconds) is on the right-hand side.}
\label{fig:languagelearning}
\end{figure*}

The greedy search strategy is necessary for \toolname to efficiently search the space of languages. However, the cost of greediness is that suboptimal grammars may be synthesized (i.e., only generating a subset of the target language), even if all selected candidates are precise. For example, consider extending the XML grammar shown in Figure~\ref{fig:xml} with the production
\begin{align*}
A_{\text{XML}}\to\textttb{<a/>}.
\end{align*}
Given the seed input
\begin{align*}
\alpha_{\text{in}}=\texttt{<a><a/></a>},
\end{align*}
phase one of \toolname synthesizes the regular expression
\begin{align*}
(\textttb{<a}(\textttb{><a/})^*\textttb{></a>})^*,
\end{align*}
which is a valid subset of $L_{\text{XML}}$. However, in phase two of \toolname, the two repetition nodes
\begin{align*}
(\textttb{><a/})^*\sss\text{and}\sss(\textttb{<a}(\textttb{><a/})^*\textttb{></a>})^*
\end{align*}
cannot be merged, since the check $\textttb{><a/}$ is invalid. Ideally, \toolname would instead synthesize the regular expression
\begin{align*}
(\textttb{<a>}(\textttb{<a/>})^*\textttb{</a>})^*,
\end{align*}
in phase one, in which case the two repetition nodes
\begin{align*}
(\textttb{<a/>})^*\sss\text{and}\sss(\textttb{<a>}(\textttb{<a/>})^*\textttb{</a>})^*
\end{align*}
are successfully merged in phase two. \toolname fails to do so because of the greedy nature of phase one. If \toolname is instead provided with the seed inputs
\begin{align*}
\{\textttb{<a/>},~\textttb{<a>hi</a>}\},
\end{align*}
then it would successfully recover the target language.

Intuitively, the greedy strategy employed by \toolname works best when the target language has fewer nondeterministic constructs (as is the case with many program input languages in practice, e.g., to ensure efficient parsing). Such grammars are less likely to have multiple incompatible candidates at each generalization step, ensuring that \toolname rarely makes suboptimal choices.

\section{Evaluation}
\label{sec:exp}

We implement our grammar synthesis algorithm in a tool called \toolname, which synthesizes a context-free grammar $\hat{C}$ given an oracle $\mathcal{O}$ and seed inputs $E_{\text{in}}\subseteq L_*$. In our first experiment, we compare \toolname to widely studied language inference algorithms, and in our second experiment, we evaluate the ability of \toolname to learn useful approximations of real program input grammars for a fuzzing client. We note that the only grammar used to guide the design our algorithm is the XML grammar, and no other grammar was used for this purpose. \toolname is implemented in Java, and all experiments are run on a 2.5 GHz Intel Core i7 CPU.

\subsection{Sampling Context-Free Grammars}
\label{sec:sample}

We describe how we randomly sample a string $\alpha$ from a context-free grammar $C$. The ability to sample implicitly defines a probability distribution $\mathcal{P}_{\mathcal{L}(C)}$ over $\mathcal{L}(C)$, which we use to measure precision and recall as in Definition~\ref{def:precisionrecall}. We also use random samples in our grammar-based fuzzer in Section~\ref{sec:expfuzz}. To describe our approach, we more generally describe how to sample $\alpha\sim\mathcal{P}_{\mathcal{L}(C,A)}$ (which is the language of strings that can be derived from nonterminal $A$ using productions in $C$). To do so, we convert the context-free grammar $C=(V,\Sigma,P,S)$ to a \emph{probabilistic context-free grammar}. For each nonterminal $A\in V$, we construct a discrete distribution $\mathcal{D}_A$ of size $|P_A|$ (where $P_A\subseteq P$ is the set of productions in $C$ for $A$). Then, we randomly sample $\alpha\sim\mathcal{P}_{\mathcal{L}(C,A)}$ as follows:
\begin{itemize}
\item Randomly sample production $(A\to A_1...A_k)\sim\mathcal{D}_A$.
\item If $A_i$ is a nonterminal, recursively sample $\alpha_i\sim\mathcal{P}_{\mathcal{L}(C,A_i)}$; otherwise, if $A_i$ is a terminal, let $\alpha_i=A_i$.
\item Return $\alpha=\alpha_1...\alpha_k$.
\end{itemize}
For simplicity, we choose $\mathcal{D}_A$ to be uniform.

\subsection{Comparison to Language Inference}
\label{sec:explearning}

\begin{figure*}
\centering
\small
\begin{tabular}{cll}
\hline
{\bf Grammar} & \multicolumn{1}{c}{{\bf Target Language $L_*$}} & \multicolumn{1}{c}{{\bf Synthesized Grammar $\hat{L}$}} \\
\hline
URL &
$\begin{array}{ll}
A\hspace{-0.1in}&\to\textttb{http}(\epsilon+\textttb{s})\textttb{:}\textttb{//}(\epsilon+\textttb{www.})[...]^*\textttb{.}[...]^*
\end{array}$
&
$\begin{array}{ll}
A\to&\hspace{-0.1in}\textttb{http://}B^*\textttb{.}C^*+\textttb{https://}B^*\textttb{.}C^* \\
&+\textttb{http://www.}B^*\textttb{.}C^*+\textttb{https://www.}B^*\textttb{.}C^* \\
B\to&\hspace{-0.1in}[...]^* \\
C\to&\hspace{-0.1in}[...]^*
\end{array}$ \\
\hline
Grep &
$\begin{array}{ll}
A\hspace{-0.1in}&\to([...]+\textttb{\char`\\(}A\textttb{\char`\\)})^*
\end{array}$
&
$\begin{array}{ll}
A\hspace{-0.1in}&\to([...]^*+((\textttb{\char`\\}(\textttb{(}A^*)^*\textttb{\char`\\})^*\textttb{)})^*)^*
\end{array}$ \\
\hline
Lisp &
$\begin{array}{ll}
A\hspace{-0.1in}&\to\textttb{(}[...][...]^*(\textvisiblespace\textvisiblespace^*([...][...]^*+A))^*\textttb{)} \\
\end{array}$
&
$\begin{array}{ll}
A\hspace{-0.1in}&\to\textttb{(}([...]^*[...]((\textvisiblespace^*\textvisiblespace A)^*\textvisiblespace^*\textvisiblespace)^*)^*[...]^*[...]\textttb{)} \\
\end{array}$ \\
\hline
XML &
$\begin{array}{ll}
A\hspace{-0.1in}&\to\textttb{<a}(\textvisiblespace\textvisiblespace^*[...][...]^*\textttb{="}[...]^*\textttb{"})^*\textttb{>}(A+[...])^*\textttb{</a>}
\end{array}$
&
$\begin{array}{ll}
A\hspace{-0.1in}&\to\textttb{<a}(\textvisiblespace^*\textvisiblespace[...]^*[...]\textttb{="}[...]^*\textttb{"})^*B^*\textttb{>}[...]^*\textttb{</a>} \\
B\hspace{-0.1in}&\to\textttb{>}[...]^*\textttb{<a}(\textvisiblespace^*\textvisiblespace[...]^*[...]\textttb{="}[...]^*\textttb{"})^*B^*\textttb{>}[...]^*\textttb{</a} \\
&\hspace{0.2in}+\textttb{>}[...]^*\textttb{<a>}[...]^*\textttb{</a} \\
\end{array}$ \\
\hline
\end{tabular}
\caption{Examples of context-free grammars that are synthesized by \toolname for the given target languages. The symbol \textvisiblespace~denotes a space. For clarity, character ranges with large numbers of characters are denoted by $[...]$.}
\label{fig:languagelearningexample}
\end{figure*}

In our first experiment, we show that \toolname can synthesize simple input grammars with much better precision and recall compared to two widely studied language inference algorithms, $L$-Star~\cite{angluin1987learning} and RPNI~\cite{oncina1992identifying}, both implemented using \texttt{libalf}~\cite{bollig2010libalf}. We also compare to a variant of \toolname with phase two omitted, which restricts \toolname to learning regular languages, which shows that the benefit of \toolname is not just its ability to synthesize non-regular properties.

\paragraph{Grammars.}
We manually wrote four grammars encoding valid inputs for various programs:
\begin{itemize}
\item A regular expression for matching URLs~\cite{stackoverflow2010url}.
\item A grammar for the regular expression accepted as input by GNU Grep~\cite{gnugrep2016grep}
\item A grammar for a simple Lisp parser~\cite{lispy2016lisp}, including support for quoted strings and comments.
\item A grammar for XML parsers~\cite{xml2008xml}, including all XML constructs (attributes, comments, CDATA sections, etc.), except that only a fixed number of tags are included (to ensure that the grammar is context-free).
\end{itemize}

\paragraph{Methods.}
For each grammar $C$, we sampled 50 seed inputs $E_{\text{in}}\subseteq L_*=\mathcal{L}(C)$ using the technique in Section~\ref{sec:sample}, and implemented an oracle $\mathcal{O}$ for $L_*$. Then, we use each algorithm to learn $L_*$ from $E_{\text{in}}$ and $\mathcal{O}$. Since the algorithms sometimes cannot scale to all 50 inputs, we incrementally give the seed inputs to the algorithms until they time out (after 300 seconds), and use the last language successfully learned without timing out.

\paragraph{$L$-Star.}
Angluin's $L$-Star algorithm learns a regular language $\hat{R}$ approximating the target language $L_*$. It takes as input a membership oracle and an \emph{equivalence oracle} $\mathcal{O}_E$; given a candidate regular language $\hat{R}$, $\mathcal{O}_E$ accepts $\hat{R}$ if $\mathcal{L}(\hat{R})=L_*$, and returns a counterexample otherwise. In our experiments, there is no way to check equivalence with the target language (i.e., the program input language). Instead, we use the variant in~\cite{angluin1987learning} where the equivalence oracle $\mathcal{O}_E$ is implemented by randomly sampling strings to search for counter-examples; we accept $\hat{R}$ if none are found after 50 samples.

\paragraph{RPNI.}
RPNI learns a regular language $\hat{R}$ given both positive examples $E_{\text{in}}$ and negative examples $E_{\text{in}}^-$. As negative examples, we sample 50 random strings not in $L_*$.

\paragraph{Results.}
We estimate the precision of $\hat{C}$ by $\frac{|E_{\text{prec}}\cap L_*|}{|E_{\text{prec}}|}$, where $E_{\text{prec}}$ consists of 1000 random samples from $\mathcal{L}(\hat{C})$, and estimate the recall of $\hat{C}$ by $\frac{|E_{\text{rec}}\cap\mathcal{L}(\hat{C})|}{|E_{\text{rec}}|}$, where $E_{\text{rec}}$ consists of 1000 random samples from $L_*$, and report the $F_1$-score $\frac{2\cdot\text{precision}\cdot\text{recall}}{\text{precision}+\text{recall}}$. The $F_1$ score is a standard metric combining precision and recall---achieving high $F_1$ score requires both high precision and high recall. We also report the running time of each algorithm, which is timed out at 300 seconds. We average all results over five runs. Figure~\ref{fig:languagelearning} shows (a) the $F_1$-score and (b) the running time of each algorithm; (c) shows how the precision, recall, and running time of \toolname vary with the number of samples in $E_{\text{in}}$.

\paragraph{Performance of \toolname.}
With just the 50 given training examples, \toolname was able to learn each grammar with an $F_1$-score of nearly $1.0$, meaning that both precision and recall were nearly 100\%. These results strongly suggest that \toolname learns most of the true structure of $L_*$. Finally, as can be seen from Figure~\ref{fig:languagelearning} (c), \toolname performs well even with few samples, and its running time likewise scales well with the number of samples. The performance of \toolname with phase two omitted (i.e., P1 in Figure~\ref{fig:languagelearning}) continues to substantially outperform $L$-Star and RPNI.

\paragraph{Phases of \toolname.}
As can be seen in Figure~\ref{fig:languagelearning} (a), \toolname consistently performs 5-10\% better than P1---i.e., the majority of the improvement of \toolname over existing algorithms is due to the active learning strategy, and the remainder is due to the ability to induce context-free grammars.

Furthermore, a consequence of our optimization when using multiple inputs (see Section~\ref{sec:multipleinputs}), \toolname is actually faster than P1---because \toolname generalizes better than P1, it uses fewer samples in $E_{\text{in}}$, thereby reducing the running time. We performed the same experiment using \toolname with the character generalization phase removed (but including both phases one and two). This variant of \toolname consistently performed similar but slightly worse than P1 both in terms of $F_1$-score and running time, so we omit results.

\paragraph{Comparison to $L$-Star and RPNI.}
$L$-Star performs well for the Grep grammar, but essentially fails to learn the other grammars, achieving either very small precision or very small recall. RPNI performs even worse, failing to learn any of the languages. $L$-Star guarantees exact learning only when a true equivalence oracle is available. Similarly, RPNI has an ``in the limit'' learning guarantee, i.e., for any enumeration of all strings $\alpha_1,\alpha_2,...\in\Sigma^*$, it eventually learns the correct language. Both of these learning guarantees require following examples:
\begin{itemize}
\item {\bf Positive:} Exercise all transitions in the minimal DFA.
\item {\bf Negative:} Reject all incorrect generalizations.
\end{itemize}
These examples are assumed to be provided either by the equivalence oracle (for $L$-Star) or in the given examples $E_{\text{in}}$ and $E_{\text{in}}^-$ (for RPNI).

However, in our setting, the equivalence oracle is unavailable to the $L$-Star algorithm and must be approximated using random sampling, so its theoretical guarantees may not hold. Indeed, random sampling rarely provides the needed examples---for example, in most runs of $L$-Star, at most two calls to the equivalence oracle found counterexamples. Similarly, for RPNI, the given examples are typically incomplete, so its theoretical guarantees likewise may not hold.

Furthermore, because these algorithms are designed to learn when the guarantees hold, they do not provide any mechanisms for recovering from failure of the assumptions, and instead fail dramatically. For example, if a terminal appears in $L_*$ but not in any seed input in $E_{\text{in}}$, then the language learned by RPNI does not contain any strings with this terminal. In contrast, \toolname incorporates generalization steps that enable it to generalize beyond behaviors in the given examples, and its carefully selected checks often provide the counterexamples needed to avoid overgeneralizing.

Additionally, while polynomial, the running times of $L$-Star and RPNI are very long. The long running time of $L$-Star is not because $L_*$ is non-regular, instead, we observe that $L$-Star algorithm issues a large number of membership queries on each of its iterations. In our setting, $L$-Star often could not even learn a four state automaton.

\paragraph{Examples.}
Figure~\ref{fig:languagelearningexample} shows examples of grammars synthesized by \toolname for the target language shown and a small set of representative seed inputs. The target languages are substantially simplified fragments of the grammars used in this experiment (to ensure clarity); the synthesized grammars are correspondingly simplified.

The structure of a synthesized grammar sometimes differs from the structure of the grammar defining the target language, even if they generate the same language. Such discrepancies occur because \toolname obtains no information about the internal representation of the target language. For example, consider the synthesized XML grammar. In a more natural grammar, the character \textttb{>} at the front of the production for $B$ would instead appear in the production for $A$, and the corresponding \textttb{>} in the production for $A$ would instead appear at the end of the production for $B$; however, this modification does not affect the generated language.

\subsection{Comparison to Fuzzers}
\label{sec:expfuzz}

\begin{figure}
\small
\centering
\begin{tabular}{c|r|r|r}
\hline
{\bf Program} & {\bf Lines of Code} & {\bf Lines in $E_{\text{in}}$} & {\bf Time (min.)} \\
\hline
sed & 2K & 3 & 0.25 \\
flex & 6K & 15 & 1.83 \\
grep & 12K & 4 & 0.17 \\
bison & 13K & 14 & 4.91 \\
xml & 123K & 7 & 2.30 \\
ruby & 120K & 80 & 229.00 \\
python & 128K & 267 & 269.00 \\
javascript & 156K & 118 & 113.00 \\
\hline
\end{tabular}
\caption{For each program, we show lines of program code, the lines of seed inputs $E_{\text{in}}$, and running time of \toolname.}
\label{fig:fuzztable}
\end{figure}

\begin{figure*}
\begin{tabular}{ccc}
\includegraphics[width=0.3\textwidth]{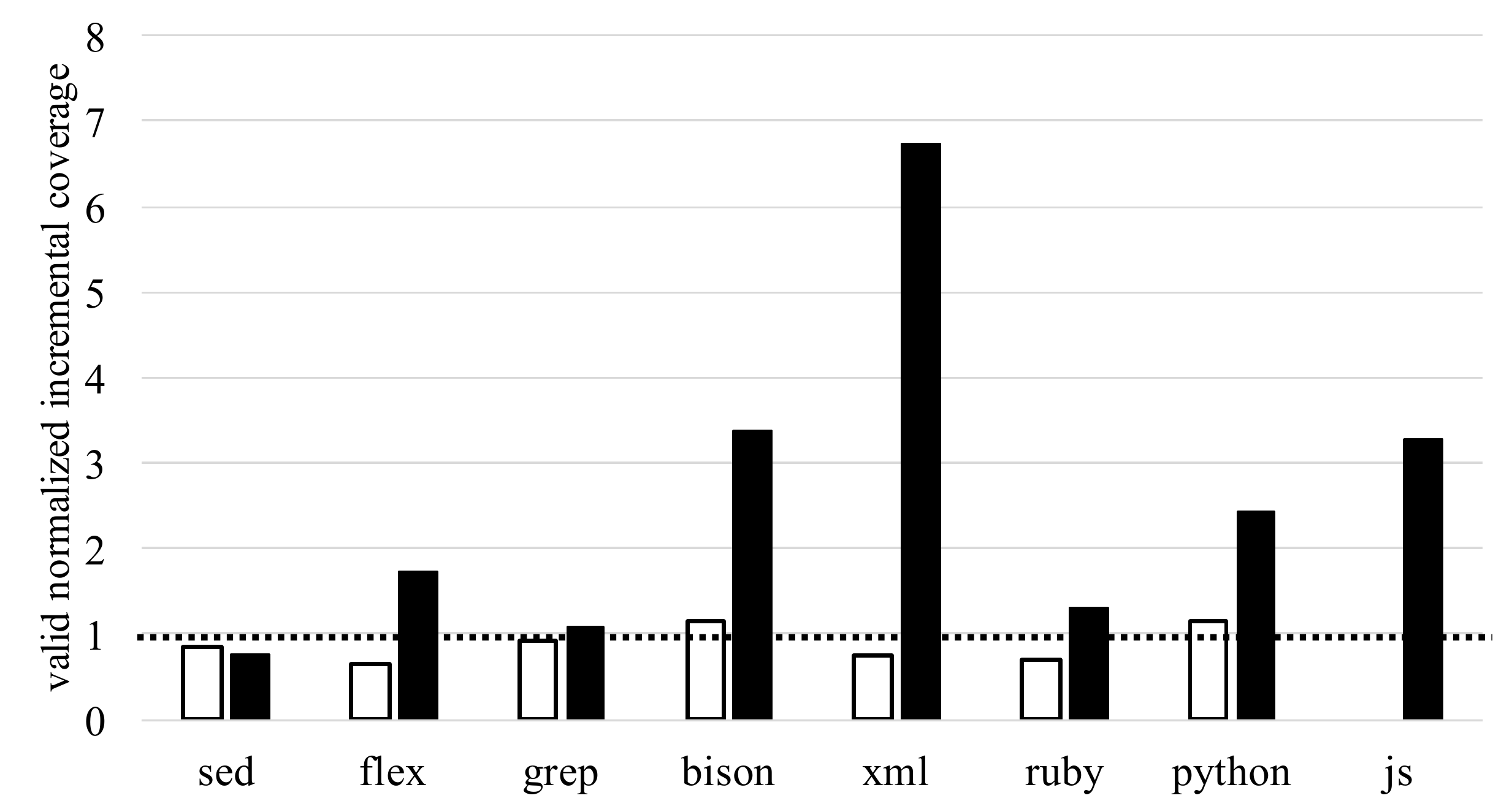}
& \includegraphics[width=0.3\textwidth]{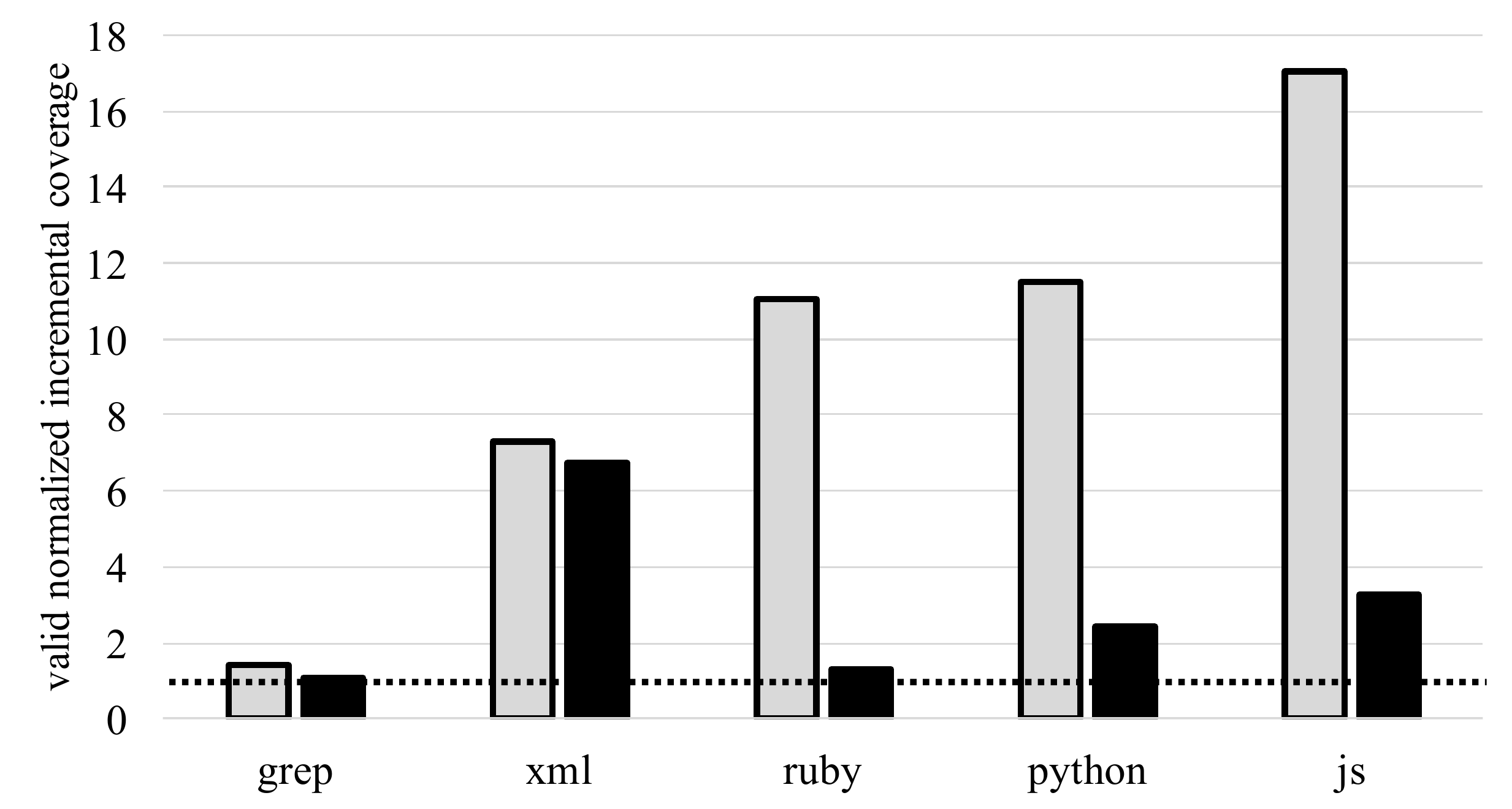}
& \includegraphics[width=0.3\textwidth]{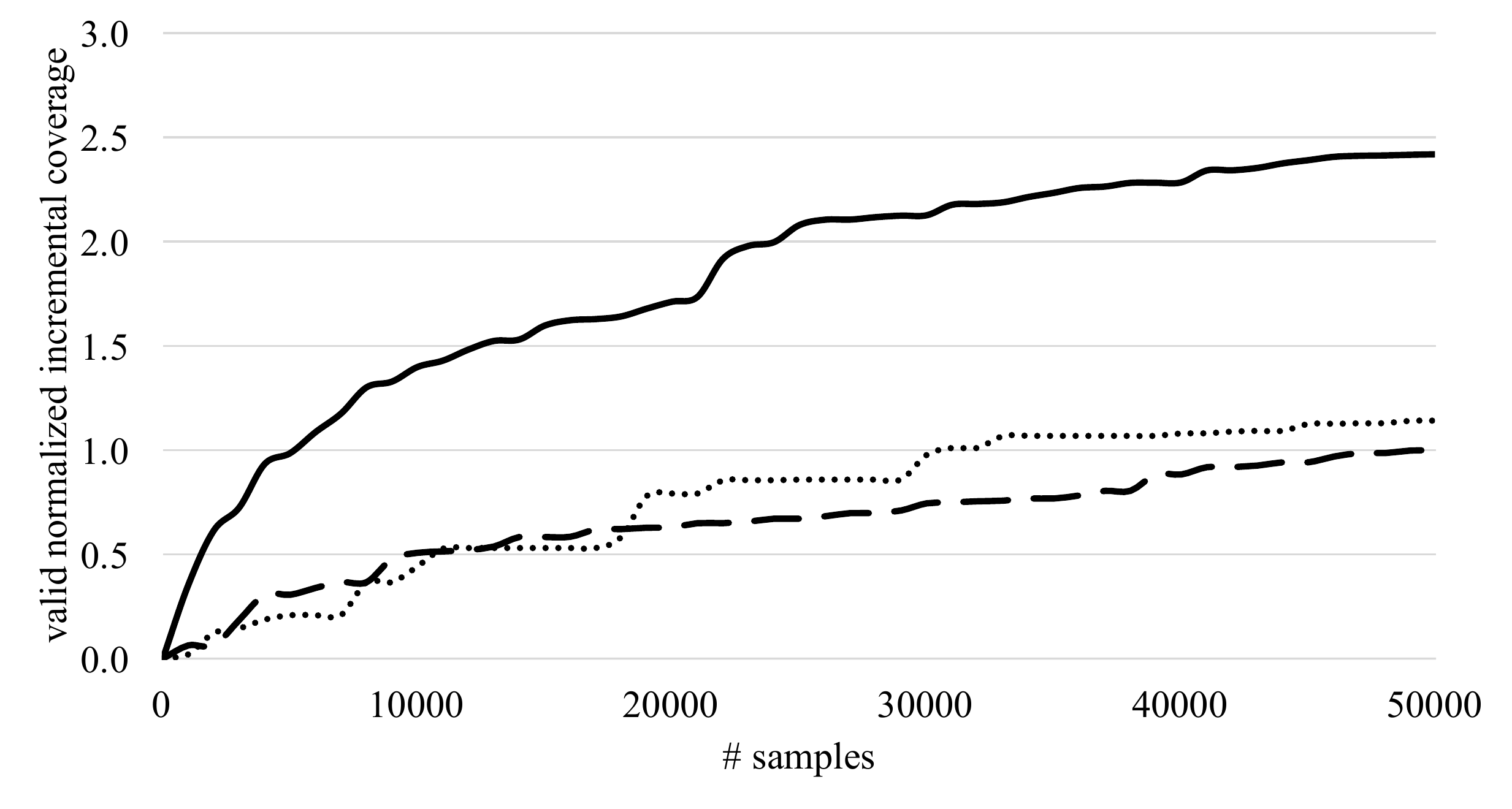} \\
(a) & (b) & (c)
\end{tabular}
\caption{In (a) we show the normalized incremental coverage restricted to valid samples for the na\"{i}ve fuzzer (black dotted line), afl-fuzz (white), and \toolname (black). In (b), we show the same metric for the na\"{i}ve fuzzer (black dotted line) and \toolname (black); grey represents either a handwritten fuzzer (for Grep and the XML parser) or a large test suite (for Python, Ruby, and Javascript). In (c), we compare the valid normalized incremental coverage of \toolname (solid) to the na\"{i}ve fuzzer (dashed) and afl-fuzz (dotted) as the number of seed inputs varies (all values are normalized by the final coverage of the na\"{i}ve fuzzer).}
\label{fig:fuzz}
\end{figure*}

For fuzzing applications such as differential testing~\cite{yang2011finding}, it is useful to obtain a large number of grammatically valid samples that exercise different functionalities of the given program. \toolname is perfectly suited to automatically generating such inputs. Given blackbox access $\mathcal{O}$ to a program with input language $L_*$ and seed inputs $E_{\text{in}}\subseteq L_*$, \toolname automatically synthesizes a context-free grammar $\hat{C}$ approximating $L_*$. Then, \toolname uses a standard grammar-based fuzzer that takes as input the synthesized grammar $\hat{C}$ and the seed inputs $E_{\text{in}}$, and randomly generates new inputs $\alpha\in\mathcal{L}(\hat{C})$ that can be used to test the program; we give details below.

In our application to fuzzing, it is acceptable for $\hat{C}$ to be an approximation---high precision suffices to ensure that most generated inputs are valid, and high recall ensures that most program behaviors have a chance of being executed.

We compare \toolname to two baseline fuzzers (described below) on the task of generating valid test inputs, and show that \toolname consistently performs significantly better.

\paragraph{Grammar-based fuzzer.}
\toolname first synthesizes a context-free grammar $\hat{C}$ approximating the target language $L_*$ of valid program inputs. Our grammar-based fuzzer, based on standard techniques~\cite{holler2012fuzzing}, takes as input the synthesized context-free grammar $\hat{C}$ and the seed inputs $E_{\text{in}}$. To generate a single random input, our grammar-based fuzzer first uniformly selects a seed input $\alpha\in E_{\text{in}}$ and constructs the parse tree for $\alpha$ according to $\hat{C}$. Second, it performs a series of $n$ modifications to $\alpha$, where $n$ is chosen uniformly between 0 and 50. A single modification is performed as follows:
\begin{itemize}
\item Randomly choose a node $N$ of the parse tree of $\alpha$.
\item Decompose $\alpha=\alpha_1\alpha_2\alpha_3$ where $\alpha_2$ is represented by the subtree with root $N$.
\item Letting $A$ be the nonterminal labeling $N$, randomly sample $\alpha'\sim\mathcal{P}_{\mathcal{L}(C,A)}$, and return $\alpha_1\alpha'\alpha_3$.
\end{itemize}

\paragraph{Afl-fuzz.}
Our first baseline fuzzer is a production fuzzer developed at Google~\cite{zalewski2015afl}, and is widely used due to its minimal setup requirements and state-of-the-art quality. It systematically modifies the input example (e.g., bit flips, copies, deletions, etc.). Unlike \toolname, afl-fuzz requires that the program be instrumented to obtain branch coverage for each execution---it uses this information to identify when an input $\alpha$ causes the program to execute new paths. It adds such inputs $\alpha$ to a worklist, and iteratively applies its fuzzing strategy to each input in the worklist. This monitoring allows it to incrementally discover deeper code paths. To run afl-fuzz on multiple inputs $E_{\text{in}}$, we fuzz each input $\alpha\in E_{\text{in}}$ in a round-robin fashion.

\paragraph{Na\"{i}ve fuzzer.}
We implement a second baseline fuzzer, which is not grammar aware. It randomly selects a seed input $\alpha\in E_{\text{in}}$ and performs $n$ random modifications to $\alpha$, where $n$ is chosen randomly between 0 and 50. A single modification of $\alpha$ consists of randomly choosing an index $i$ in $\alpha=\sigma_1...\sigma_k$, and either deleting the terminal $\sigma_i$ or inserting a randomly chosen terminal $\sigma\in\Sigma$ before $\sigma_i$.

\paragraph{Programs.}
We set up each fuzzer on eight programs that include front-ends of language interpreters (Python, Ruby, and Mozilla's Javascript engine SpiderMonkey), Unix utilities that take structured inputs (Grep, Sed, Flex, and Bison), and an XML parser. We were unable to setup afl-fuzz for Javascript, showing that even production fuzzers can have setup difficulties when they require code instrumentation. For interpreters (e.g., the Python interpreter), we focus on fuzzing just the parser (e.g., the Python parser) since the input grammar of the interpreter contains elements such as variable and function names, use-before-define errors, etc., that are out of scope for our grammar synthesis algorithm. To fuzz the parser, we ``wrap'' the input inside a conditional statement, which ensures that the input is never executed. For example, we convert the Python input (\texttt{print `hi'}) to the input (\texttt{if False: print `hi'}). Then, syntactically incorrect inputs are rejected, but inputs that are syntactically correct but possibly have runtime errors are accepted.

\paragraph{Seed inputs.}
To fuzz a program, we use a small number of seed inputs $E_{\text{in}}\subseteq L_*$ that capture interesting semantics of the target language $L_*$. These seed inputs were obtained either from documentation and tutorials or from small test suites that came with the program.

\paragraph{Methods.}
Coverage is difficult to interpret because a large amount of code in each program is unreachable due to configuration, test code that cannot be executed, and other unused functionality. Therefore, we use a relative measure of coverage to evaluate performance. As before, all results are averaged over five runs.

For each program and fuzzer, we generate 50,000 samples $E\subseteq\Sigma^*$ by running the fuzzer on the program. First, we restrict $E$ to valid inputs, i.e., $E\cap L_*$. In particular, the \emph{valid coverage} of $E$, computed using gcov, is
\begin{align*}
\frac{\#(\text{lines covered by }E\cap L_*)}{\#(\text{lines coverable})}.
\end{align*}
Next, the \emph{valid incremental coverage} of $E$ is the percentage of code covered by valid inputs in $E$, ignoring those already covered by the seed inputs $E_{\text{in}}$ (thereby measuring the ability to discover inputs that execute new code paths):
\begin{align*}
\frac{\#(\text{lines covered by }E\cap L_*\text{ but not covered by }E_{\text{in}})}{\#(\text{lines coverable but not covered by }E_{\text{in}})}.
\end{align*}
Finally, to enable comparison across programs, the \emph{valid normalized incremental coverage} normalizes the incremental coverage by a baseline $E_{\text{base}}$:
\begin{align*}
\frac{\text{valid incremental coverage of }E}{\text{valid incremental coverage of }E_{\text{base}}}.
\end{align*}
In particular, we use samples from the na\"{i}ve fuzzer as $E_{\text{base}}$.

\paragraph{Results.}
In Figure~\ref{fig:fuzztable}, we show various statistics for the eight programs we use and for the corresponding seed inputs $E_{\text{in}}$. We also show the time \toolname needed to synthesize an approximation of the program input grammar. In Figure~\ref{fig:fuzz} (a), we show the valid normalized incremental coverages of the various fuzzers. In (b), for five of our programs, we show a proxy for the ``upper bound'' in coverage that is achievable---for Grep and the XML parser, we show the valid normalized incremental coverage achieved by our handwritten grammars, and for Python, Ruby, and Javascript, we show the valid normalized incremental coverage of a large test suite (each more than 100,000 lines of code). In (c), we show how coverage varies with the number of samples for Python.

\paragraph{Comparison to baselines.}
As can be seen from Figure~\ref{fig:fuzz} (a), \toolname (black) is effective at generating valid inputs that exercise new code paths, significantly outperforming both the na\"{i}ve fuzzer (black dotted line) and afl-fuzz (white) except on Grep (where it only performs slightly better) and Sed (where it actually performs slightly worse). Since these programs have a relatively simple input format, using a grammar-based fuzzer is understandably less effective. For the remaining six programs, our grammar-based fuzzer performs between 1.3 and 7 times better than the na\"{i}ve fuzzer.

\paragraph{Comparison to proxy for the upper bound.}
Figure~\ref{fig:fuzz} (b) compares \toolname (black bars) to a proxy for the upper bound of coverage, i.e., handwritten grammars or large test suites (grey bars). For Grep, both \toolname and the na\"{i}ve fuzzer achieve coverage close to the handwritten grammar. For the XML parser, \toolname significantly outperforms the na\"{i}ve fuzzer, achieving coverage close to the handwritten grammar. For Python and Javascript, \toolname is able to recover a significantly larger fraction of the upper bound compared to the na\"{i}ve fuzzer. However, a sizable gap remains, which is expected since the test suites are very large (each having at least 100,000 lines of code) and are specifically designed to test the respective programs. We provided fewer seed inputs for Ruby, which explains why \toolname outperformed the na\"{i}ve fuzzer by a smaller amount (about 30\%).

\paragraph{Coverage over time.}
Figure~\ref{fig:fuzz} (c) shows how the valid normalized incremental coverage varies with the number of samples. \toolname (solid) quickly finds a number of high-coverage inputs that the other fuzzers cannot, and continues to find more inputs that execute new lines of code.

\paragraph{Examples.}
The synthesized grammars are too large to show. Instead, as an example, a fragment of the synthesized XML grammar is
\begin{align*}
A\to&~\textttb{<a}\textvisiblespace^*\textvisiblespace[...]^*[...]\textttb{="}[...]^*\textttb{"}B^*\textttb{>}[...]^*\textttb{</a>} \\
B\to&~\textttb{>}[...]^*\textttb{<a}\textvisiblespace^*\textvisiblespace[...]^*[...]\textttb{="}[...]^*\textttb{"}B^*\textttb{>}[...]^*\textttb{</a} \\
&+\textttb{>}[...]^*\textttb{<a>}[...]^*\textttb{</a}.
\end{align*}
This grammar is identical to the synthesized XML grammar shown in Figure~\ref{fig:languagelearningexample}, except that attributes cannot be repeated. In particular, \toolname learns that attributes cannot be repeated since XML semantics requires that different attributes have different names---for example, the input string \texttt{<a a="" a=""></a>} is invalid. Therefore, repeating the attribute would lead to overgeneralization, so this construct is rejected by \toolname. Indeed, this constraint on attribute names is not a context-free property, so as expected, \toolname learns a subset of the XML input language.

Figure~\ref{fig:fuzzexample} shows an example of a valid sample from the grammar synthesized by \toolname for the XML parser. As can be seen, the sample contains many XML constructs, including nested tags, attributes, comments, and processing instructions.

\begin{figure}
\small
\begin{verbatim}
<a>
  \%
  <a QE="{>_">
    C
    <a    _="#">
      ">q(+_[s:?>^0+
      <a  _eD="{@">
        :"<a>. q</a>1+%
      </a>
      y<!--       y-->y
    </a>
    _<a>x</a>y
  </a>
  xy<?q  xy?>xy<?xV <?By_![?>x
</a>
\end{verbatim}
\caption{An example of a valid sample from the grammar synthesized by \toolname for the XML parser. For clarity, the string has been formatted with additional whitespace.}
\label{fig:fuzzexample}
\end{figure}

\section{Related Work}

\paragraph{Mining input formats.}
The work most closely related to our own is~\cite{hoschele2016mining}, which uses dynamic taint analysis to trace the flow of each input character, and uses this information to reconstruct the input grammar. More broadly, there has been work on reverse engineering network protocol message formats~\cite{caballero2007polyglot,wondracek2008automatic,lin2008deriving,lin2010reverse}, though these papers focus on learning and understanding the structure of given inputs rather than learning a grammar; for example,~\cite{caballero2007polyglot} looks for variables representing the internal parser state to determine the protocol, and~\cite{lin2008deriving} constructs syntax trees for given inputs. All of these techniques rely on static and dynamic analysis methods intended to reverse engineer parsers of specific designs.

In contrast, our approach is fully blackbox and depends only on the language accepted by the program, not the specific design of the program's parser. In addition, our approach can be used when the program cannot be instrumented, for instance, to learn the input format for a remote program. Finally, the programs we consider have more complex input formats than most previously examined programs.

\paragraph{Learning theory.}
There has been a line of work in learning theory (often referred to as \emph{grammar induction} or \emph{grammar inference}) aiming to learn a grammar from either examples or oracles (or both); see~\cite{de2010grammatical} for a survey. The most well known algorithms are $L$-Star~\cite{angluin1987learning} and RPNI~\cite{oncina1992identifying}. These algorithms have a number of applications including model checking~\cite{giannakopoulou2012symbolic}, model-assisted fuzzing~\cite{cho2011mace,choi2013guided}, verification~\cite{vardhan2004learning}, and specification inference~\cite{botinvcan2013sigma}. To the best of our knowledge, our work is the first to focus on the application of learning common program input languages from positive examples and membership oracles.

Additionally,~\cite{lee1996learning} discusses approaches to learning context-free grammars, including from positive examples and a membership oracle. As they discuss, these algorithms are often either slow~\cite{solomonoff1959new} or do not generalize well~\cite{knobe1976method}.

\paragraph{Bayesian language learning.}
A related line of work aims to learn probabilistic grammars from examples alone~\cite{stolcke1994inducing,stolcke1994bayesian}. These algorithms study a different setting than ours, in particular, they are given access to positive (and sometimes negative) examples, but do not assume access to a membership oracle. These algorithms typically identify frequently occurring patterns that are likely to correspond to nonterminals in the grammar. More precisely, these algorithms are typically Bayesian learning algorithms that operate by putting a prior over the space of grammars, and then computing the most likely grammar conditioned on the given examples. To achieve statistically significant results, these algorithms require a large number of input examples.

In contrast, our algorithm leverages access to the membership oracle, enabling it to use actively generated examples to determine which patterns are actually in the grammar. Therefore, our algorithm works well even when only a few seed inputs are available. While it may be possible to modify existing Bayesian language learning algorithms to fit this setting, to the best of our knowledge, no such active learning variants of these algorithms have been proposed.

Additionally, whereas this literature aims to learn a probabilistic grammar, our grammar synthesis algorithm learns a deterministic grammar. The difference is how we measure approximation quality---in particular, even though our definitions of precision and recall require distributions over $L_*$ and $\hat{L}$, they still measure the approximation quality of $\hat{L}$ deterministically, i.e., the predicates $\alpha\in L_*$ and $\alpha\in\hat{L}$ are binary rather than probabilistic.

\paragraph{Blackbox fuzzing.}
Numerous approaches to automated test generation have been proposed; we refer to~\cite{anand2013orchestrated} for a survey. Approaches to fuzzing (i.e., random test case generation) broadly fall into two categories: whitebox (i.e., statically inspect the program to guide test generation) and blackbox (i.e., rely only on concrete program executions). Blackbox fuzzing has been used to test software for several decades; for example,~\cite{sauder1962general} randomly tests COBOL compilers and~\cite{purdom1972sentence} generated random inputs to test parsers. An early application of blackbox fuzzing to find bugs in real-world programs was~\cite{miller1990empirical}, who executed Unix utilities on random byte sequences to discover crashing inputs. Subsequently, there have been many approaches using blackbox fuzzing with dynamic analysis to find bugs and security vulnerabilities~\cite{forrester2000empirical,sutton2005art,miller2006empirical}; see~\cite{sutton2007fuzzing} for a survey. Finally, afl-fuzz~\cite{zalewski2015afl} is almost blackbox, requiring only simple instrumentation to guide the search.

\paragraph{Whitebox fuzzing.}
Approaches to whitebox fuzzing~\cite{godefroid2008automated,artzi2008finding} typically build on \emph{dynamic symbolic execution}~\cite{godefroid2005dart,sen2005cute,cadar2008exe,cadar2008klee,cadar2013symbolic}; given a concrete input example, these approaches use a combination of symbolic execution and dynamic execution to construct a constraint system whose solutions are inputs that execute new program branches compared to the given input. It can be challenging to scale these approaches to large programs~\cite{ganesh2009taint}. Therefore, approaches relying on more imprecise input have been studied; for example, taint analysis~\cite{ganesh2009taint}, or extracting specific information such as a checksum computation~\cite{wang2010taintscope}.

\paragraph{Grammar-based fuzzing.}
Many fuzzing approaches leverage a user-defined grammar to generate valid inputs, which can greatly increase coverage. For example, blackbox fuzzing has been combined with manually written grammars to test compilers~\cite{lindig2005random,yang2011finding}; see~\cite{boujarwah1997compiler} for a survey. Such techniques have also been used to fuzz interpreters; for example,~\cite{holler2012fuzzing} develops a framework for grammar-based testing and applies it to find bugs in both Javascript and PHP interpreters.

Grammar-based approaches have also been used in conjunction with whitebox techniques. For example,~\cite{godefroid2008grammar} fuzzes a just-in-time compiler for Javascript using a handwritten Javascript grammar in conjunction with a technique for solving constraints over grammars, and~\cite{majumdar2007directed} combines exhaustive enumeration of valid inputs with symbolic execution techniques to improve coverage. In ~\cite{sutton2007fuzzing}, Chapter 21 gives a case study developing a grammar for the Adobe Flash file format. Our approach can complement existing grammar-based fuzzers by automatically generating a grammar.

Finally, there has been some work on inferring grammars for fuzzing~\cite{viide2008experiences}, but focusing on simple languages such as compression formats. To the best of our knowledge, our work is the first targeted at learning complex program input languages that contain recursive structure, e.g., XML, regular expression formats, and programming language syntax.

\paragraph{Synthesis.}
Finally, our approach uses machinery related to some of the recent work on programming by example---in particular, a systematic search guided by a meta-grammar. This approach has been used to synthesize string~\cite{gulwani2011automating}, number~\cite{singh2012synthesizing}, and table~\cite{harris2011spreadsheet} transformations (and combinations thereof~\cite{perelman2014test,polozov2015flashmeta}), as well as recursive programs~\cite{feser2015synthesizing,albarghouthi2013recursive} and parsers~\cite{leung2015interactive}. Unlike these approaches, our approach exploits an oracle to reject invalid candidates.

\section{Conclusion}

We have presented \toolname, the first practical algorithm for inferring program input grammars, and demonstrated its value in an application to fuzz testing. We believe \toolname may be valuable beyond fuzzing, e.g., to generate whitelists of inputs or to reverse engineer input formats.

\acks

{\small This material is based on research sponsored by DARPA under agreement number FA84750-14-2-0006.  The U.S. Government is authorized to reproduce and distribute reprints for Governmental purposes notwithstanding any copyright notation thereon.  The views and conclusions herein are those of the authors and should not be interpreted as necessarily representing the official policies or endorsements either expressed or implied of DARPA or the U.S. Government.  This work was also supported by NSF grant CCF-1160904 and a Google Fellowship.}

%Acknowledgments, if needed.

% We recommend abbrvnat bibliography style.

\bibliographystyle{abbrvnat}

% The bibliography should be embedded for final submission.

%\begin{thebibliography}{}
%\softraggedright
\softraggedright
\balance
\bibliography{paper}
%\end{thebibliography}

\clearpage

\appendix
\section{Properties of Phase One}

We prove the desired properties discussed in Section~\ref{sec:algo} for the generalization steps proposed in Section~\ref{sec:phaseone}. First, we prove Proposition~\ref{prop:phaseonecandidate}, which says that the candidates in phase one are monotone. Next, we prove Proposition~\ref{prop:context}, which says that the contexts constructed by phase one satisfy (\ref{eqn:context}); as discussed in Section~\ref{sec:phaseonecheck}, this result implies that the corresponding checks $\alpha$ constructed in phase one are valid (i.e., $\alpha\in\tilde{L}\setminus\hat{L}_i$). 

\subsection{Proof of Proposition~\ref{prop:phaseonecandidate}}
\label{sec:phaseonecandidateproof}

There are two cases:
\begin{itemize}
\item {\bf Repetitions:} Every candidate has form (omitting bracketed substrings) $R'=P\blue{\alpha_1}\blue{\alpha}_{\blue{2}}^*\blue{\alpha_3}Q$, where the current language is $R=P\blue{\alpha}Q$ and $\alpha=\alpha_1\alpha_2\alpha_3$. Since $\alpha\in\mathcal{L}(\blue{\alpha_1}\blue{\alpha}_{\blue{2}}^*\blue{\alpha_3})$, it is clear that
\begin{align*}
\mathcal{L}(R)=\mathcal{L}(P\blue{\alpha}Q)\subseteq\mathcal{L}(P\blue{\alpha_1}\blue{\alpha}_{\blue{2}}^*\blue{\alpha_3}Q)=\mathcal{L}(R').
\end{align*}
\item {\bf Alternations:} Every candidate has form (omitting bracketed substrings) $R'=P(\blue{\alpha_1}+\blue{\alpha_2})Q$, where the current language is $R=P\blue{\alpha}Q$ and $\alpha=\alpha_1\alpha_2$. Note that an bracketed expression $[\alpha]_{\text{alt}}$ always occurs within a repetition, so the candidate has form
\begin{align*}
R'&=...(...+(\blue{\alpha_1}+\blue{\alpha_2})+...)^*... \\
&=...(...+(\blue{\alpha_1}+\blue{\alpha_2})^*+...)^*...,
\end{align*}
so since $\alpha\in(\alpha_1+\alpha_2)^*$, we have
\begin{align*}
\mathcal{L}(R)=\mathcal{L}(P\blue{\alpha}Q)\subseteq\mathcal{L}(P(\blue{\alpha_1}+\blue{\alpha_2})^*Q)=\mathcal{L}(R').
\end{align*}
\end{itemize}
The result follows.~$\square$

\subsection{Proof of Proposition~\ref{prop:context}}
\label{sec:contextproof}

We prove by induction. The initial context $(\epsilon,\epsilon)$ for $[\alpha_{\text{in}}]_{\text{rep}}$ clearly satisfies (\ref{eqn:context}). Next, assume that the context $(\gamma,\delta)$ for the current language satisfies (\ref{eqn:context}). There are two cases:
\begin{itemize}
\item {\bf Repetitions:} Suppose that the current language is $R=P[\blue{\alpha}]_{\text{rep}}Q$ and the candidate is $R'=P\blue{\alpha_1}([\blue{\alpha_2}]_{\text{alt}})^*[\blue{\alpha_3}]_{\text{alt}}$. Then, the context constructed for $[\blue{\alpha_2}]_{\text{alt}}$ is $(\gamma',\delta')=(\gamma\alpha_1,\alpha_3\delta)$. Also, let $P'=$~``$P\blue{\alpha_1}($'' and $Q'=$~``$)^*\blue{\alpha_3}Q$'', so $R'=P'\blue{\alpha_2}Q'$. Then, for any $\alpha'\in\Sigma^*$, we have
\begin{align*}
\gamma'\alpha'\delta'=\gamma\alpha_1\alpha'\alpha_3\delta
&\in\mathcal{L}(P\blue{\alpha_1}\blue{\alpha'}\blue{\alpha_3}Q) \\
&\subseteq\mathcal{L}(P\blue{\alpha_1}(\blue{\alpha'})^*\blue{\alpha_3}Q) \\
&=\mathcal{L}(P'\blue{\alpha'}Q'),
\end{align*}
where the first inclusion follows by applying (\ref{eqn:context}) to the context $(\gamma,\delta)$ with $\alpha_1\alpha'\alpha_3\in\Sigma^*$. Therefore, the context $(\gamma',\delta')$ satisfies (\ref{eqn:context}). Similarly, the context constructed for $[\blue{\alpha_3}]_{\text{rep}}$ is $(\gamma',\delta')=(\gamma\alpha_1\alpha_2,\delta)$. Also, let $P'=P\blue{\alpha_1}\blue{\alpha}_{\blue{2}}^*$ and $Q'=Q$, so $R'=P'\blue{\alpha_3}Q'$. Then, for any $\alpha'\in\Sigma^*$, we have
\begin{align*}
\gamma'\alpha'\delta'=\gamma\alpha_1\alpha_2\alpha'
&\in\mathcal{L}(P\blue{\alpha_1\alpha_2\alpha'}Q) \\
&\subseteq\mathcal{L}(P\blue{\alpha_1\alpha_2^*\alpha'}Q) \\
&=\mathcal{L}(P'\blue{\alpha'}Q'),
\end{align*}
where the first inclusion follows by applying (\ref{eqn:context}) to the context $(\gamma,\delta)$ with $\alpha_1\alpha_2\alpha'\in\Sigma^*$. Therefore, the context $(\gamma',\delta')$ satisfies (\ref{eqn:context}).
\item {\bf Alterations:} Suppose that the current language is $R=P[\blue{\alpha}]_{\text{alt}}Q$ and the candidate is $R'=P([\blue{\alpha_1}_{\text{rep}}+[\blue{\alpha_2}]_{\text{alt}})Q$. Then, the context constructed for $[\blue{\alpha_1}]_{\text{rep}}$ is $(\gamma',\delta')=(\gamma,\alpha_2\delta)$. Also, let $P'=$~``$P($'' and $Q'=$~``$+\blue{\alpha_2})Q$'', so $R'=P'\blue{\alpha_2}Q'$. Then, for any $\alpha'\in\Sigma^*$, we have
\begin{align*}
\gamma'\alpha'\delta'=\gamma\alpha'\alpha_2\delta
&\in\mathcal{L}(P\blue{\alpha'\alpha_2}Q) \\
&=\mathcal{L}(P(\blue{\alpha'}+\blue{\alpha_2})^*Q) \\
&=\mathcal{L}(P(\blue{\alpha'}+\blue{\alpha_2})Q) \\
&=\mathcal{L}(P'\blue{\alpha'}Q'),
\end{align*}
where the inclusion follows by applying (\ref{eqn:context}) to the context $(\gamma,\delta)$ with $\alpha'\alpha_2\in\Sigma^*$, and the equality on the third line follows as in the proof of Proposition~\ref{prop:phaseonecandidate} (in Section~\ref{sec:phaseonecandidateproof}).
\end{itemize}
The claim follows.~$\square$

\section{Expressiveness of Phase One}

In this section, we prove the expressiveness results discussed in Section~\ref{sec:phaseonecandidate}.

\subsection{Correspondence to Derivations in $\mathcal{C}_{\text{regex}}$}
\label{sec:correspondenceproof}

In this section, we prove Proposition~\ref{prop:correspondence}, which says that derivations in $\mathcal{C}_{\text{regex}}$ can be transformed to series of generalization steps in phase one of our algorithm. In particular, consider the derivation of a regular expression $R\in\mathcal{L}(\mathcal{C}_{\text{regex}})$:
\begin{align*}
T_{\text{rep}}=\eta_1\Rightarrow...\Rightarrow\eta_n=R.
\end{align*}
We prove that for each $i$, there is a series of generalization steps
\begin{align*}
R_i\Rightarrow R_{i+1}\Rightarrow...\Rightarrow R_n=R
\end{align*}
such that each $R_j$ (for $i\le j\le n$) maps to $\eta_j$ in the way defined in Section~\ref{sec:phaseonecandidate} (i.e., by replacing $[\blue{\alpha}]_{\tau}$ with $T_{\tau}$); we express this mapping as $\eta_j=\overline{R_j}$. The result follows since for $i=1$, we get $[\blue{\alpha}]_{\text{rep}}=R_1\Rightarrow...\Rightarrow R_n=R$, so we can take $\alpha_{\text{in}}=\alpha$.

We prove by (backward) induction on the derivation. The base case $i=n$ is trivial, since $\eta_n\in\mathcal{L}(\mathcal{C}_{\text{regex}})$, so we can take $R_n=\eta_n$ since $\overline{R_n}=R_n=\eta_n$. Now, suppose that we have a series of generalization steps $R_{i+1}\Rightarrow...\Rightarrow R_n=R$ that satisfies the claimed property. It suffices to show that we can construct $R_i$ such that $R_i\Rightarrow R_{i+1}$ is a generalization step and $\overline{R_i}=\eta_i$. Consider the following cases for the step $\eta_i\Rightarrow\eta_{i+1}$ in the derivation:
\begin{itemize}
\item Step $\mu T_{\text{rep}}\nu\Rightarrow\mu\beta T_{\text{alt}}^*T_{\text{rep}}\nu$: Then, we must have
\begin{align*}
R_{i+1}=P\blue{\alpha_1}[\blue{\alpha_2}]_{\text{alt}}[\alpha_3]_{\text{rep}}Q,
\end{align*}
where $\overline{P}=\mu$, $\overline{Q}=\nu$, and $\alpha_1=\beta$. Also, since $R_{i+1}$ is valid, we have $\alpha_1,\alpha_2,\alpha_3\not=\epsilon$. Therefore, we can take
\begin{align*}
R_i=P[\blue{\alpha}]_{\text{rep}}Q,
\end{align*}
where $\alpha=\alpha_1\alpha_2\alpha_3\not=\epsilon$. The remaining productions for $T_{\text{rep}}$ are similar. In particular, the assumption that $\beta\not=\epsilon$ in these derivations is needed to ensure that $\alpha\not=\epsilon$.
\item Step $\mu T_{\text{alt}}\nu\Rightarrow\mu(T_{\text{rep}}+T_{\text{alt}})\nu$: Then, we must have
\begin{align*}
R_{i+1}=P([\blue{\alpha_1}]_{\text{rep}}+[\blue{\alpha_2}]_{\text{alt}})Q,
\end{align*}
where $\overline{P}=\mu$ and $\overline{Q}=\nu$. Also, since $R_{i+1}$ is valid, we have $\alpha_1,\alpha_2\not=\epsilon$. Therefore, we can take
\begin{align*}
R_i=P[\blue{\alpha}]_{\text{alt}}Q,
\end{align*}
where $\alpha=\alpha_1\alpha_2\not=\epsilon$. The remaining production for $T_{\text{alt}}$ is similar.
\end{itemize}
The result follows.~$\square$

\subsection{Expressiveness of $\mathcal{C}_{\text{regex}}$}
\label{sec:metagrammarproof}

In this section, we prove Proposition~\ref{prop:metagrammar}, which says that any regular language can be expressed as $\mathcal{L}(R_1+...+R_m)$, where $R_1,...,R_m\in\mathcal{L}(\mathcal{C}_{\text{regex}})$ are regular expressions that can be synthesized by phase one of our algorithm.

We slightly modify $\mathcal{C}_{\text{regex}}$, by introducing a new nonterminal $T_{\text{regex}}$, taking $T_{\text{regex}}$ to be the start symbol, and adding productions
\begin{align*}
T_{\text{regex}}&::=\bar{\epsilon}\mid T_{\text{alt}}\mid\bar{\epsilon}+T_{\text{alt}},
\end{align*}
where $\bar{\epsilon}\in\Sigma_{\text{regex}}$ is a newly introduced terminal denoting the regular expression for the empty language. This modification has two effects:
\begin{itemize}
\item Now, regular expressions $R\in\mathcal{L}(\mathcal{C}_{\text{regex}})$ can have top-level alternations.
\item Furthermore, the top-level alternation can explicitly include the empty string $\bar{\epsilon}$ (e.g., $R=\bar{\epsilon}+\textttb{a}$).
\end{itemize}
As described in Section~\ref{sec:phaseonecandidate}, the first modification can be addressed by using multiple inputs (see Section~\ref{sec:multipleinputs}), which allows our algorithm to learn top-level alternations. The second modification can be addressed by including a seed input $\bar{\epsilon}\in E_{\text{in}}$, in which case phase one of our algorithm synthesizes $\bar{\epsilon}$ (since there is nothing for it to generalize).

Now, let the context-free grammar $\tilde{C}_{\text{regex}}$ be a standard grammar for regular expressions:
\begin{align}
\label{eqn:regex}
T::=\beta\mid TT\mid T+T\mid T^*.
\end{align}

It suffices to show that for any $R\in\mathcal{L}(\mathcal{C}_{\text{regex}})$, there exists $R'\in\mathcal{L}(\tilde{C}_{\text{regex}})$ such that $\mathcal{L}(R)=\mathcal{L}(R')$ (which we express as $R\equiv R'$).

First, we prove the result for $\mathcal{C}_{\text{regex}}^{\epsilon}$, which is identical to $\mathcal{C}_{\text{regex}}$ except that we allow $\beta=\epsilon$. Let $R\in \mathcal{L}(\tilde{C}_{\text{regex}})$. Suppose that either $R=S_1+S_2$, $R=S_1S_2$, or $R=\beta$. We claim that we can express $R$ as
\begin{align*}
\addtocounter{equation}{1}\tag{\theequation}\label{eqn:canonical}
R&\equiv X_1+...+X_n \\
X_i&=Y_{i,1}...Y_{i,k_i}\hspace{0.25in}(1\le i\le n)
\end{align*}
where either $Y_{i,j}=\beta$ or $Y_{i,j}=W_{i,j}^*$ for each $i$ and $j$. Consider two possibilities:
\begin{itemize}
\item Suppose $R$ can be expressed in the form (\ref{eqn:canonical}), but $Y_{i,j}=Z_1+Z_2$. Then
\begin{align*}
X_i&=Y_{i,1}...Y_{i,j}...Y_{i,k_i} \\
&=Y_{i,1}...(Z_1+Z_2)...Y_{i,k_i} \\
&\equiv Y_{i,1}...Z_1...Y_{i,k_i}+Y_{i,1}...Z_2...Y_{i,k_i}
\end{align*}
which is again in the form (\ref{eqn:canonical}).
\item Suppose $R$ has the form (\ref{eqn:canonical}), but $Y_{i,j}=Z_1Z_2$. Then
\begin{align*}
X_i=Y_{i,1}...Y_{i,j}...Y_{i,k_i}=Y_{i,1}...Z_1Z_2...Y_{i,k_i}
\end{align*}
which is again in the form (\ref{eqn:canonical}).
\end{itemize}
Note that either $R=S_1+S_2$ or $R=S_1S_2$, so $R$ starts in the form (\ref{eqn:canonical}). Therefore, we can repeatedly apply the above two transformations until $Y_{i,j}=\beta$ or $Y_{i,j}=W_{i,j}^*$ for every $i$ and $j$. This process must terminate because the parse tree for $R$ is finite, so the claim follows.

Now, we construct $R'\in \mathcal{L}(\mathcal{C}_{\text{regex}}^{\epsilon},T_{\text{alt}})$ such that $R\equiv R'$ by structural induction. First, suppose that either $R=S_1+S_2$, $R=S_1S_2$, or $R=\beta$. Then we can express $R$ in the form (\ref{eqn:canonical}). By induction, $W_{i,j}\equiv W_{i,j}'$ for some $W_{i,j}'\in \mathcal{L}(\mathcal{C}_{\text{regex}}^{\epsilon},T_{\text{alt}})$ for every $i$ and $j$. By the definition of $T_{\text{rep}}$, we have $X_i\in \mathcal{L}(\mathcal{C}_{\text{regex}}^{\epsilon},T_{\text{rep}})$, so by the definition of $T_{\text{alt}}$, we have $R\in \mathcal{L}(\mathcal{C}_{\text{regex}}^{\epsilon},T_{\text{alt}})$, so the inductive step follows.

Alternatively, suppose $R=S^*$. If $S=S_1^*$, then $R\equiv S_1^*$, so without loss of generality assume $S=S_1+S_2$, $S=S_1S_2$, or $S=\beta$, so by the previous argument, we have $S\in \mathcal{L}(\mathcal{C}_{\text{regex}}^{\epsilon},T_{\text{alt}})$. Since $T_{\text{alt}}::=T_{\text{rep}}$ and $T_{\text{rep}}::=T_{\text{alt}}^*$, we have $R\in \mathcal{L}(\mathcal{C}_{\text{regex}}^{\epsilon},T_{\text{alt}})$, so again the inductive step follows. Finally, since $T::=T_{\text{alt}}$, we have $R\in \mathcal{L}(\mathcal{C}_{\text{regex}}^{\epsilon})$.

Now, we modify the above proof to show that as long as $\epsilon\not\in \mathcal{L}(R)$, we have $R\in \mathcal{L}(\mathcal{C}_{\text{regex}},T_{\text{alt}})$. As before, we proceed by structural induction. Suppose that either $R=S_1+S_2$, $R=S_1S_2$, or $R=\beta$, so we can express $R$ in the form (\ref{eqn:canonical}). First, consider the case $Y_{i,j}=\beta$; if $\beta=\epsilon$, we can remove $Y_{i,j}$ from $X_i$ unless $k_i=1$. However, if $Y_{i,j}=\beta=\epsilon$ and $k_i=1$, whence $X_i=\epsilon$ so $\epsilon\in \mathcal{L}(R)$, a contradiction; hence, we can always drop $Y_{i,j}$ such that $Y_{i,j}=\epsilon$. For the remaining $Y_{i,j}=\beta$, we have $Y_{i,j}\in \mathcal{L}(\mathcal{C}_{\text{regex}},T_{\text{rep}})$ by the definition of $\mathcal{C}_{\text{regex}}$.

Second, consider the case $Y_{i,j}=Z_{i,j}^*$. Let $Z_{i,j}'$ be a regular expression such that $\mathcal{L}(Z_{i,j}')=\mathcal{L}(Z_{i,j})-\{\epsilon\}$, and note that
\begin{align*}
Y_{i,j}=Z_{i,j}^*\equiv(Z_{i,j}')^*.
\end{align*}
By induction, we know that $Z_{i,j}\in \mathcal{L}(\mathcal{C}_{\text{regex}},T_{\text{alt}})$, so $Y_{i,j}'=(Z_{i,j}')^*\in \mathcal{L}(\mathcal{C}_{\text{regex}},T_{\text{rep}})$ by the definition of $\mathcal{C}_{\text{regex}}$.

For each $X_i$, we remove every $Y_{i,j}=\beta=\epsilon$ and replace every $Y_{i,j}=Z_{i,j}^*$ with $Y_{i,j}'=(Z_{i,j}')^*$ to produce $X_i'\equiv X_i$. By definition of $\mathcal{C}_{\text{regex}}$, we have $X_i\in \mathcal{L}(\mathcal{C}_{\text{regex}},T_{\text{rep}})$, so $R\in \mathcal{L}(\mathcal{C}_{\text{regex}},T_{\text{alt}})$ as claimed; now, the case $R=S^*$ follows by the same argument as before.

For any $R$ such that $\epsilon\in \mathcal{L}(R)$, we can write $R=\epsilon+S$ where $\epsilon\not\in \mathcal{L}(S)$ and apply the above argument to $S$. Since $T::=\epsilon+T_{\text{alt}}$ is a production in $\mathcal{C}_{\text{regex}}$, we have shown that $R\in \mathcal{L}(\mathcal{C}_{\text{regex}})$ for any regular expression $R$.~$\square$

\section{Properties of Phase Two}

We prove the desired properties discussed in Section~\ref{sec:algo} for the generalization steps proposed in Section~\ref{sec:phasetwo}. As discussed in Section~\ref{sec:phasetwocandidate}, the candidates constructed in phase two are clearly monotone (since equating nonterminals in a context-free grammar can only enlarge the generated language). We prove Proposition~\ref{prop:merge}, which formalizes our intuition about how candidates constructed in phase two merge repetition subexpressions; as discussed in Section~\ref{sec:phasetwocheck}, this result implies that the checks constructed in phase two are valid.

\subsection{Proof of Proposition~\ref{prop:merge}}
\label{sec:propmergeproof}

In this section, we sketch a proof of Proposition~\ref{prop:merge}. In particular, we show that if we merge two nonterminals $(A_i',A_j')\in M$ by equating them in the context-free grammar $\hat{C}$ (translated from $\hat{R}$) to obtain $\tilde{C}$, then the repetition subexpressions $R$ in $\hat{R}=PRQ$ (corresponding to $A_i'$) and $R'$ in $\hat{R}=P'R'Q'$ (corresponding to $A_j'$) are merged; i.e., $\mathcal{L}(PR'Q)\subseteq\mathcal{L}(\tilde{C})$ and $\mathcal{L}(P'RQ')\subseteq\mathcal{L}(\tilde{C})$. While we prove the result for the translation $\hat{C}$ of $\hat{R}$, note that (i) subsequent merges can only enlarge the generated language, and (ii) the order in which merges are performed does not affect the final context-free grammar, so the result holds for any step of phase two of our algorithm.

Note that equating two nonterminals $(A_i',A_j')\in M$ in $\hat{C}$ is equivalent to adding productions $A_i'\to A_j'$ and $A_j'\to A_i'$ to $\hat{C}$. Therefore, Proposition~\ref{prop:merge} shows that both $\mathcal{L}(PR'Q)\subseteq\mathcal{L}(\tilde{C})$ and $\mathcal{L}(P'RQ')\subseteq\mathcal{L}(\tilde{C})$. It suffices to show that adding $A_i'\to A_j'$ to $\hat{C}$ results in the context-free grammar $\tilde{C}$ satisfying $\mathcal{L}(PR'Q)\subseteq\mathcal{L}(\tilde{C})$ (intuitively, this is a one-sided merge that only merges $\hat{R}'$ into $\hat{R}$, not vice versa).

We use the fact that our algorithm for translating a regular expression to a context-free grammars works more generally for any regular expression $R\in\mathcal{L}(\mathcal{C}_{\text{regex}})$ derived from $T_{\text{rep}}$ in according to the meta-grammar $\mathcal{C}_{\text{regex}}$. In particular, if we consider the series of generalization steps
\begin{align*}
\alpha_{\text{in}}=R_1\Rightarrow...\Rightarrow R_n=\hat{R},
\end{align*}
we get a corresponding derivation
\begin{align*}
T_{\text{rep}}^{(1)}=\eta_1\Rightarrow...\Rightarrow\eta_n=\hat{R}
\end{align*}
in $\mathcal{C}_{\text{regex}}$ as described in Section~\ref{sec:phaseonecandidate}. Similarly to the labels on bracketed strings in the series of generalization steps, we label each nonterminal in the derivation with the index at which it is expanded. For example, for the derivation corresponding to the the series of generalization steps in Figure~\ref{fig:exampletranslation} is
\begin{align*}
&T_{\text{rep}}^{(1)} \\
&\Rightarrow(T_{\text{alt}}^{(2)})^* \\
&\Rightarrow(T_{\text{rep}}^{(3)})^* \\
&\Rightarrow(\textttb{<a>}(T_{\text{alt}}^{(5)})^*T_{\text{rep}}^{(4)})^* \\
&\Rightarrow (\textttb{<a>}(T_{\text{alt}}^{(5)})^*\textttb{</a>})^* \\
&\Rightarrow (\textttb{<a>}(T_{\text{rep}}^{(8)}+T_{\text{alt}}^{(6)})^*\textttb{</a>})^* \\
&\Rightarrow (\textttb{<a>}(T_{\text{rep}}^{(8)}+T_{\text{rep}}^{(7)})^*\textttb{</a>})^* \\
&\Rightarrow (\textttb{<a>}(T_{\text{rep}}^{(8)}+\textttb{i})^*\textttb{</a>})^* \\
&\Rightarrow (\textttb{<a>}(\textttb{h}+\textttb{i})^*\textttb{</a>})^*
\end{align*}
Now, each nonterminal $A_i$ is associated to step $i$ in the derivation, and we add productions for $A_i$ depending on step $i$ in the derivation (and auxiliary nonterminals $A_i'$ if step $i$ in the derivation expands nonterminal $T_{\text{rep}}$ in the meta-grammar):
\begin{itemize}
\item Step $\mu T_{\text{rep}}^{(i)}\nu\Rightarrow\mu\beta(T_{\text{alt}}^{(j)})^*T_{\text{rep}}^{(k)}\nu$: We add productions $A_i\to\beta A_i'A_k$ and $A_i'\to\epsilon\mid A_i'A_j$.
\item Step $\mu T_{\text{alt}}^{(i)}\nu\Rightarrow\mu(T_{\text{rep}}^{(j)}+T_{\text{alt}}^{(k)})\nu$: We add production $A_i\to A_j\mid A_k$.
\end{itemize}
Now, consider step $i$ in the derivation, where productions for $A_i$ and $A_i'$ were added to $\hat{C}$. Then, step $i$ of the derivation has form
\begin{align*}
\mu T_{\text{rep}}^{(i)}\nu\Rightarrow\mu\beta(T_{\text{alt}}^{(j)})^*T_{\text{rep}}^{(k)}\nu.
\end{align*}
We can assume without loss of generality that we expand $T_{\text{rep}}^{(i)}$ last; i.e., $\mu=\overline{\mu}=P$ and $\nu=\overline{\nu}=Q$ do not contain any nonterminals. Therefore, the derivation has form
\begin{align*}
(\eta_1=T_{\text{rep}}^{(1)})
&\Rightarrow... \\
&\Rightarrow(\eta_i=PT_{\text{rep}}^{(i)}Q) \\
&\Rightarrow(\eta_{i+1}=P\beta(T_{\text{alt}}^{(j)})^*T_{\text{rep}}^{(k)}Q) \\
&\Rightarrow... \\
&\Rightarrow(\eta_n=PRQ).
\end{align*}
Now, note that the following derivation is also in $\mathcal{C}_{\text{regex}}$:
\begin{align*}
(\eta_1=T_{\text{rep}}^{(1)})
&\Rightarrow... \\
&\Rightarrow(\eta_i=PT_{\text{rep}}^{(i)}Q) \\
&\Rightarrow(\eta_{i+1}'=P\beta'(T_{\text{alt}}^{(j')})^*T_{\text{rep}}^{(k')}Q) \\
&\Rightarrow... \\
&\Rightarrow\eta_{n'}'=PR'Q
\end{align*}
since $R'$ can be derived from $T_{\text{rep}}$. Note that $\hat{R}'=PR'Q$ is exactly the regular expression produced by this derivation. Then, let $\hat{C}'$ be the context-free grammar obtained by applying our translation algorithm to $\hat{R}'$ using this derivation.

Note that $\hat{C}'$ has the same productions as $\hat{C}$, except the productions for $A_i$ in $\hat{C}$ (i.e., all productions added on step $i$ of the derivation and after) have been replaced with productions $A_i$ in $\hat{C}'$ such that $\mathcal{L}(\hat{C}',A_i)=\mathcal{L}(R')$. Since $\mathcal{L}(R')\subseteq\mathcal{L}(\tilde{C},A_i)$, and the nonterminals involved in the productions for $A_i$ do not occur in $\tilde{C}$, it is clear that adding the productions for $A_i$ in $\hat{C}'$ to $\tilde{C}$ does not modify $\mathcal{L}(\tilde{C})$. By construction, the other productions in $\hat{C}'$ are in $\hat{C}$, so they are also in $\tilde{C}$. Therefore, $\mathcal{L}(\hat{C}')\subseteq\mathcal{L}(\tilde{C})$. The result follows, since $\mathcal{L}(\hat{C}')=\mathcal{L}(\hat{R}')=\mathcal{L}(PR'Q)$.~$\square$

\section{Expressiveness of Phase Two}
\label{sec:phasetwoexpressivenessproof}

In this section, we give a proof sketch of the expressiveness result stated in Proposition~\ref{prop:phasetwoexpressiveness} of Section~\ref{sec:phasetwoexpressiveness}. Let $C$ be a generalized matching parentheses grammar. Suppose that nonterminal $S_i$ ($1\le i\le n$) corresponds to production
\begin{align*}
S_i\to R_i(S_{i_1}+...+S_{i_{k_i}})^*R_i'.
\end{align*}
First, we need to identify a context such that $S_i$ can occur in a derivation in $C$; in particular, we want to construct a derivation of the form
\begin{align*}
S_0=S_{i,1}
&\Rightarrow R_{i,1}S_{i,2}R_{i,1}' \\
&\Rightarrow R_{i,1}R_{i,2}S_{i,3}R_{i,2}'R_{i,1}' \\
&\Rightarrow... \\
&\Rightarrow R_{i,1}...R_{i,h_i}S_iR_{i,h_i}'...R_{i,1}'.
\end{align*}
To do so, we construct a directed graph $G$ with vertices $\{S_1,...,S_n\}$ and edges $S_i\to S_j$ whenever the production for $S_i$ has form
\begin{align*}
S_i\to R_i(...+S_j+...)^*R_i'.
\end{align*}
In other words, an edge indicates that $S_j$ is contained in a derivation of $S_i$. Then, we can constructed the desired derivation using a spanning tree rooted at $S_1$, in particular, by examining the path
\begin{align*}
S_1=S_{i,1}\to...\to S_{h_i}\to S_i
\end{align*}
from $S_1$ to $S_i$ in this spanning tree. Note that if no path exists, then $S_i$ cannot occur in any derivation of $S_1$.

Now, for each pair of regular expressions $R_i$ and $R_i'$ ($1\le i\le n$), let $\alpha_i\in\mathcal{L}(R_iR_i')\subseteq\mathcal{L}(C,S_i)$. Then, let
\begin{align*}
X_i&=R_{i,1}...R_{i,h_i}Y_iR_{i,h_i}'...R_{i,1}' \\
Y_i&=(R_i(\alpha_{i_1}^*+...+\alpha_{i_{k_i}}^*)R_i')^*.
\end{align*}
Intuitively, $X_i$ is constructed according to the derivation of $S_1$ containing $S_i$, and $Y_i$ is constructed using the production for $S_i$. In paricular, by construction, $\mathcal{L}(X_i)\subseteq\mathcal{L}(C)$.

Consider the following regular expression:
\begin{align*}
X&=X_1+...+X_n \\
M&=\{(Y_i,\alpha_{j_k}^*)\mid i=j_k\}.
\end{align*}

We claim that translating $X$ and $M$ into a context-free grammar yields a grammar $C'$ such that $\mathcal{L}(C)=\mathcal{L}(C')$. First, we show that each production in $C$ is also in $C'$, which implies that $\mathcal{L}(C)\subseteq\mathcal{L}(C')$. In particular, note that the translation algorithm introduces exactly one nonterminal for each $Y_i$, since two repetition nodes $Y_i$ and $Y_j$ are never merged together, and every other repetition node in $X$ is merged with a $Y_i$ node. Let $S_i'$ be the nonterminal introduced for $Y_i$; since each $\alpha_{i_j}$ is merged with $Y_{i_j}$, the production added to $C'$ is
\begin{align*}
S_i'\to(R_i(S_{i_1}'+...+S_{i_{k_i}}')^*R_i')^*,
\end{align*}
which is equivalent to the production for $S_i$ in $C$.

Next, we show that $\mathcal{L}(X)\subseteq\mathcal{L}(C)$. First, note that by construction, $\mathcal{L}(X_i)\subseteq\mathcal{L}(C)$ for each $1\le i\le n$, so $\mathcal{L}(X)\subseteq\mathcal{L}(C)$. Second, applying each merge in $M$ does not affect this invariant, since $Y_i$ and $\alpha_{j_k}^*$ can both be replaced with $S_i=S_{j_k}$. Therefore, $\mathcal{L}(C)=\mathcal{L}(C')$.~$\square$

\end{document}